# Formation of Zn and Pb sulfides in a redox-sensitive modern system due to high atmospheric fallout

Beata Smieja-Król [a,*], Mirosława Pawlyta [b], Mariola Kądziołka-Gaweł [c], Barbara Fiałkiewicz-Kozieł [d]

[a] Institute of Earth Sciences, Faculty of Natural Sciences, University of Silesia in Katowice, Będzińska 60, Sosnowiec, Poland

[b] Institute of Engineering Materials and Biomaterials, Faculty of Mechanical Engineering, Silesian University of Technology, 18A Konarskiego Street, Gliwice, 44-100, Poland

[c] Institute of Physics, University of Silesia in Katowice, 75 Pułku Piechoty 1, Chorzów, Poland

[d] Faculty of Geographical and Geological Sciences, Adam Mickiewicz University, Krygowskiego 10, Poznań, Poland

Human activities have led to a considerable increase in trace metal cycling in recent times. The mobilized elements get subjected to a variety of processes leading eventually to their re-concentration. The polluted sites, transformed by Earth's surface processes, become similar to metal accumulations known from geological records. This study examines authigenic metal sulfide mineralization in two peatlands polluted by atmospheric deposition from a nearby Pb-Zn smelter. We use the polluted peatlands as a small-scale model of Zn-Cd-Pb sulfide deposit to determine the role of organic matter in ore genesis and the textural and structural organization of biogenic precipitates.

The study shows that the air-derived metal enrichment (up to 2.3 g Zn kg$^{-1}$, 1.1 g Pb kg$^{-1}$, and 62 mg Cd kg$^{-1}$) is retained in a thin layer (~30 cm) around 10-15 cm below the peat surface. A combination of focused ion beam (FIB) technology and scanning (SEM) and transmission (TEM) electron microscopy reveals that micrometric spheroids are most characteristic for ZnS and (Zn,Cd)S, although the sulfides readily form pseudomorphs after different plant tissues resulting in much larger aggregates. The aggregates have a complex polycrystalline sphalerite structure much more advanced than typically obtained during low-temperature synthesis or observed in other modern occurrences. Platy highly-disordered radially-aggregated submicrometre crystals develop within the time constraints of several decades in the cold (~15°C) and acid (pH 3.4-4.4) peat. The less abundant Pb sulfides occur as small cube-like crystals (<1µm) between ZnS or as flat irregular or square patches on plant root macrofossils. All PbS are crystalline and defect-free. Pb ion complexation with dissolved and solid organic matter is probably responsible for the low number and equilibrium shape of PbS crystals. Iron is absent in the authigenic sulfide mineralization and occurs entirely as organically bound ferric iron (Fe$^{3+}$), as revealed by Mössbauer spectroscopy. The different affinity of metals to organic matter enhances the precipitation of Zn and Cd as sulfides over Pb and Fe. Our findings demonstrate that human activities lead to the formation of near-surface stratiform metal sulfide accumulations in peat, and the polluted sites can be of use to understand and reconstruct ancient ore deposits' genesis and mechanisms of formation.

# 1. Introduction

Sulfide precipitation is an important process in metals cycling, known from recent and past environments (e.g., Farquhar et al., 2010; Rickard et al., 2017). As most metal sulfides have very low solubilities, the sulfides can precipitate even from dilute solutions (Labrenz et al., 2000; Kosolapov et al., 2004). The only requirement is the availability of sulfide ions. In temperatures <~80°C, the sulfide ions originate from microbial sulfate reduction (MSR) in the presence of abundant biodegradable organic matter. Due to high activation energy, which is overcome by enzymatic processes in MSR, the abiotic reaction is kinetically inhibited in this temperature range (Cross et al., 2004). The sulfide formation is dominated by ferrous sulfides (pyrite, mackinawite, greigite) because of the overwhelming abundance of Fe relative to other chalcophile metals (Rickard et al., 2017); unique geochemical conditions are required for the precipitation of the less common sulfides in the low-temperature range.

A few nonferrous sulfide accumulations are known from modern environments, all in terrestrial settings (Lett and Fletcher, 1980; Lee et al., 1984; Yoon et al. 2012; Awid-Pascual et al. 2015; Knappova et al., 2019). Awid-Pascual et al. (2015) reported a Zn-Pb Grieves Siding prospect site (Tasmania), hosted by Quaternary peat of a thickness between <1 and 20m, composed of organic remnants derived from woody parts of terrestrial plants. The peat contains up to 28.6 wt% Zn and up to 3.8 wt% Pb. SEM investigations confirmed extensive authigenic mineralization containing Zn, S, O, Al, Pb, and Si. Mixed and intermediate sulfur valences, including sulfides and oxysulfides, are suggested to coexist in the colloform precipitates. Mineralized microorganisms of the same composition were also found (Awid-Pascual et al., 2015). A naturally zinc-enriched (up to 7.1% Zn) peatland with authigenic ZnS mineralization with mixed sphalerite and wurtzite structures was described by Martinez et al. (2007) and Yoon et al. (2012).

In addition to the formation of metal sulfides driven by nature alone, human activities contribute significantly to the process by increasing the Earth's surface reservoir of trace elements since the onset of the Industrial Revolution. Ore smelting/refining, fossil fuel combustion, and vehicle emissions lead to large-scale dispersion of the elements previously stored in ancient ore deposits (Nriagu and Pacyna, 1988; Pacyna and Pacyna, 2001). The mobilized elements tend to translocate and can re-concentrate in surface systems where appropriate conditions exist. A concept of human-made deposits has already been introduced by Saryg-ool et al. (2017). Zinc, Cd, Cu, Pb, and Hg sulfides were confirmed to precipitate at contaminated sites rich in natural organic matter. These include freshwater canal and river sediments, wetlands, soils, and flooded mines (Sobolewski, 1996; Gammons and Frandsen, 2001; Large et al., 2001; Sonke et al., 2002; Moreau et al., 2004; Smieja-Król et al. 2015; Myagkaya et al., 2016; Ciszewski et al., 2017; Lazareva et al., 2019; Mantha et al., 2019; Quevedo et al., 2020). Experimental studies confirm the importance of organic matter in facilitating metal sulfide formation and persistence in water fluctuating systems (Lau and Hsu-Kim, 2008; Weber et al., 2009; Hofacker et al., 2013; Hoffmann et al., 2020).

Peatlands cover 2–3% of the Earth's land and are vast global repositories of partially decomposed organic detritus that accumulate under waterlogged conditions by the constant addition of organic material (Clymo 1983; Gorham, 1991). Peat organic matter is porous (surface area> 200$m^2$/g), reactive, i.e., prone to microbial degradation, and has high sorption/complexation potential toward dissolved solids (Asplund et al., 1976; Brown et al., 2000). Consequently, peat deposits located in the vicinity of industrial centers are shown to accumulate high concentrations of metal(loid)s in the top layers (e.g., Jones and Hao 1993, Nieminen et al., 2002; Sonke et al., 2002; Mihaljevič et al., 2006; Linton et al., 2007; Souter and Watmough, 2016). Sonke et al. (2002) measured concentrations up to 4.7 wt.% Zn, 1.1 wt.% Pb and 0.1 wt.% Cd within an upper 30 cm layer of a mire located close to a zinc smelter complex in Belgium. A rapid diagenetic transformation of dust-deposited metal oxides into sulfides is indicated (Sonke et al., 2002).

This study focuses on metal retention and accumulation, accompanied by the precipitation of metal sulfides, in two peatlands enriched in sulfur and trace metals from atmospheric fallout of Zn-Pb smelter emission. We hypothesize that polluted peatlands can be used as a small-scale model of Zn-Cd-Pb sulfide deposit of biogenic origin. There is a growing body of evidence that organic matter and microbial sulfidogenesis matter in ore formation over geological times (e.g., Large et al., 1998; Bawden et al. 2003; Taylor, 2004; Tornos et al., 2008; Barrie et al., 2009; Herlec et al., 2010). However, little is known about the influence of microorganisms and the biodegraded organic matter on the mineralization process and metals deliverance to the crystalization site. Most features from the early stage of sulfide mineral precipitation are lost or overprinted by later processes. Still, the close association of microorganisms with sulfide minerals is documented by Kucha (1988), Kucha et al. (1990), Kucha et al. (2010) for various ancient deposits.

The study aims to reveal the role of organic matter in the mineralization process and the textural and structural development of the precipitates at short-time scales, i.e., not burdened with subsequent geological processes. The nano- to micrometer-scale of metal sulfide organization is revealed using a combination of focused ion beam (FIB) technology with scanning (SEM) and transmission electron microscopy (TEM).

The peatlands development, plant/macrofossil composition, trophy status, time of metal deposition, and the mineral constituents of dust and peat are well characterized (Smieja-Król et al., 2010; Fiałkiewicz-Kozieł et al., 2014; Smieja-Król and Bauerek, 2015).

## 2. Material and methods

### 2.1. Site description

The peatlands are located in Upper Silesia Upland, within a forested inland dune field formed during Pleistocene in a flat depression of soft Triassic shales and sandstones bordered to the north and south by Triassic limestones (Fiałkiewicz-Kozieł et al., 2014). The dune field constitutes the southern-most part of the so-called 'European sand belt' formed along the Weichselian ice fronts throughout northwestern and central Europe (Jankowski, 2012; Zeeberg, 1998). Numerous peat deposits fill deflation hollows reflecting local sand relocation during dune formation (Szczypek, 1977). Today, most are lost because of long-time drainage and forest management in the region. The investigated peatlands lay ca 17 km apart and are relatively well preserved, although both show amelioration attempts in the past, probably in the 18/19$^{th}$ century (Fiałkiewicz-Kozieł et al., 2014). Both are supplied mainly by rainwater; the concentration of Mg, Ca, Na, and K ions in porewater is comparable to the precipitation in the region. However, some groundwater inflow from the surrounding fluvioglacial sands to the lower peat layers is probable (Smieja-Król and Bauerek, 2015). Bagno Bruch peatland (BB: N50°31′, E19° 2′; 300 m a.s.l.) covers a total area of about 39 ha. The peat thickness is 90 cm on average but reaches a maximum depth of 205 cm in its northern part. The BB peatland experiences dynamic variations in the water-table level. The water table can be as low as −60 cm in dry seasons and level with the peatland surface during winter and springtime (Smieja-Król et al., 2010). Bagno Mikołeska (BM; N50°33′38″, E18°49′2″; 268 m a.s.l.) is a small (c. 5 ha) peatland elongated W-E in a depression bordered to the north, north-west, and north-east by a large forested dune complex. The peat thickness exceeds 200 cm. The water-table level stays close to the peatland surface throughout the year.

The plant composition classifies both peatlands as poor fens. Due to a variety of rare and endangered species (e.g., *Andromeda polifolia*, *Eriophorum vaginatum*, *Ledum palustre*, *Oxycoccus palustris*, *Drosera rotundifolia*, and a variety of *Sphagnum* species), BB wetland is protected within the Natura 2000 network under the European Union Habitats Directive (PLH240035), and BM peatland has a more local level of formal protection (ecological site). A detailed description of the peatlands is given in Fiałkiewicz-Kozieł et al. (2014).

Previous studies indicate elevated concentrations of several trace metals in the upper peat layers due to atmospheric fallout in both dissolved and particulate form (Smieja-Król et al., 2010; Smieja-Król and Fiałkiewicz-Kozieł, 2014; Smieja-Król and Bauerek, 2015). The environment has been impacted by the Upper Silesia industrial region located directly south of the peatlands (Fig. S1). Early Cretaceous Zn-Pb-Fe ore-bearing dolomites of the Mississippi Valley Type (MVT) deposits (Heijlen et al., 2003) have been exploited since medieval times. Bituminous coal from the Carboniferous Upper Silesian Coal Basin has been in use since the 16th century (Smieja-Król et al., 2019). Two smelters are the most crucial metal and sulfur sources located closest to the peatlands (Fig. S1). Fryderyk Pb smelter in Strzybnica operated in 1786–1933 with unknown emission for that time, but the production was 0.1 Mg Ag, 225,900 Mg Pb, and 19,500 Mg Pb oxide, respectively, with a maximum in 1905 (Jaros, 1969). Zn-Pb smelter in Miasteczko Śl. (1968-present) uses the Imperial Smelting Process for ore processing and emitted 5,288 Mg Zn, 2,836 Mg Pb, 50 Mg Cd, and 117,000 Mg $SO_2$, respectively, the first 25 years of operation. After that time, the emission dropped significantly (Fig. S2).

### 2.2. Sampling and analyses of solids

Two cores were collected from each peatland using a stainless steel 10 × 10 $cm^2$ Wardenaar corer (Wardenaar, 1986). The BM1 and BM2 cores were taken from the central part of BM peatland, ~100 m apart. BB1 and BB2 cores were collected from the northern, thickest part of BB peatland, ca 40 m apart. In the laboratory, the cores were sectioned into 1 cm slices with a stainless-steel knife. The uppermost part (0-7cm) was cut into three slices due to the low weight of the material.

500 mg of every second core slice was air-dried, homogenized, re-dried at 105 °C, and ashed at 460 °C for 24 h before extraction in concentrated 65% HNO3 (100 °C; 1 h). Concentrations of Zn, Pb, Cd, and As were determined using a PerkinElmer ICP-MS spectrometer (Elan 6100 DRC-e; Institute of Environmental Engineering, Polish Academy of Sciences). Rhodium was used as an internal spike during the ICP-MS analyses. Iron concentration was determined using an atomic absorption spectrometer (AAS, Thermo Scientific iCE 3000 series) in six samples collected separately to conduct $^{57}$Fe Mössbauer spectroscopy (see below). All ion concentrations were corrected for procedural blanks. The reference material NIMT/UOE/FM/001 (Yafa et al., 2004) was used to assess the degree of the metal(loid)s recovery (for details, see Table S1). Abundances of total sulfur (TS) were determined in duplicate samples using an Eltra CS-500 IR analyzer. Eltra standard was used for calibration and quality control (Table S1).

Fragments of peat core slices for scanning electron microscopy (SEM) were wet stored at 4°C, then fixed with 2% glutaraldehyde for 1–2 h, dehydrated through a series of ethanol (EtOH) washes (15 min at 50, 75, 96, and 3×100 %), air-dried, mounted on aluminum specimen stubs, and sputter-coated with gold. Additionally, some samples were air-dried and, without any additional treatment, carbon-coated prior to examination. Samples were examined using two scanning electron microscopes, each coupled to an energy-dispersive X-ray analyzer system (EDS), namely, SEM-FEI Quanta 250 and a high-resolution FEI Inspect F SEM. Accelerating voltage of 5–25 kV and working distances between 5 and 10 mm were used to obtain EDS microanalyses, backscattered electron (BSE), and secondary-electron (SE) images.

Samples for the $^{57}$Fe Mössbauer spectroscopy and transmission electron microscopy (TEM) investigations were taken separately from the peatlands to limit air penetration. A pit was cut using stainless steel knife within ~1m distance from the BM1 and BB1 peat core collection sites, respectively. Peat samples were taken using head-cut syringe samplers directly from the pit wall to argon-filled conical glass flasks. The flasks were immediately refilled with argon and tightly closed.

Samples for TEM analyses were dried in a vacuum dryer at room temperature. Macrofossils of plant roots were mounted on aluminum specimen stubs and carbon-coated. Microsites with ZnS mineralization were selected using the SEM-BSE detector. Two specimens (10 x 10µm x 10-100 nm) were prepared by Focused Ion Beam (FIB) technique using the SEM/Ga-FIB Helios NanoLab™ 600i microscope (Fig. S3). The thin sections were kept under an inert atmosphere during storage and transportation. The time between the specimens preparation and TEM observation was 16 h. TEM investigations were performed on an aberration-corrected FEI Titan electron microscope (S/TEM Titan 80–300 from FEI Co.) operating at 300 kV. The chemical composition was determined in the same apparatus using energy dispersive spectroscopy (EDS). Bright-field (BF) and high-angle annular dark-field (HAADF) detectors were used in STEM imaging. Bright-field (BF), dark-field (DF), and high-resolution (HR) modes were used in the TEM system.

The $^{57}$Fe Mössbauer spectroscopy was conducted consecutively on wet samples shortly after their collection. The samples from deeper peat layers were analyzed first. $^{57}$Fe Mössbauer transmission spectra were recorded in vacuum at room temperature with an MS96 Mössbauer spectrometer and a linear arrangement of a $^{57}$Co: Rh source, a multichannel analyzer with 1024 channels (before folding), an absorber, and a detector. The spectrometer was calibrated at room temperature with a 30 μm thick α-Fe foil. A numerical analysis of the Mössbauer spectra was performed using the WMOSS program (author I. Prisecaru). Spectra were fitted as a superposition of several doublets. The decomposition into doublets was performed by a Lorentzian function.

**2.3. Sampling and chemical analyses of pore water**

Two piezometers (PVC, diameter 50 mm) were installed close to each peat core collection site to determine the pore water properties of the peatlands at two depth ranges. A short piezometer, perforated at a depth interval of 5-40 cm, was used to monitor the water table dynamic and aqueous chemistry of the subsurface, polluted peat layer. The second, one-meter long, with a perforation between 60-100 cm below the peat surface, was used to investigate the influence of groundwater. Water measurements were conducted five times during the frost-free season in 2010 and 2011. Before each sampling event, the piezometers were purged of old water and allowed to equilibrate before sample collection. Electrical conductivity (EC), pH, redox potential (Eh), and dissolved oxygen were measured directly in the field using an integrated WTW MultiLine P4 meter. Field Eh values were corrected to the standard hydrogen electrode. Samples for water analyses were collected in acid-washed polyethylene bottles and immediately stored in a cold box for transport to the laboratory. Water for sulfide analyses was collected to glass flasks with stoppers. The position of the water table relative to the peat surface was measured.

Metals and As were determined using the PerkinElmer ICP-MS spectrometer and $Cl^-$, $SO_4^{2-}$ using a Metrohm ion chromatograph after filtering the samples through 0.45 membrane filters the day after collection. Concentrations of sulfides were determined by the thiomercurimetric method based on the titration with o-hydroxymercuribenzoic acid (HMB) in the presence of dithizone as an indicator and sodium hydroxide and disodium EDTA as a buffer to red color (Wronski, 1971). Dissolved organic carbon (DOC) was determined in unacidified water samples filtered through 0.45 membrane filters using a Total Organic Carbon Analyzer (Shimadzu TOC 5000). The PHREEQC geochemical code (Parkhurst and Appelo, 2013) was used to calculate Eh values from redox couples of the porewater.

## 3. Results

### 3.1. Porewater properties

The porewater measured at the four sites and two depth intervals has a relatively similar composition (Table 1 and Table S2), suggesting no significant spatial and temporal variations in peatland's water properties. The porewater composition is also comparable to other poor fens of temperate climate (Shotyk, 1988). The shallow porewater (Table 1) is acidic (pH: 3.4-4.4) and poor in conductive ions (EC: 34-142 $\mu$S cm$^{-1}$). Accordingly, the concentration of trace metals is low (Table 1), generally reflecting that of precipitation in the region (Smieja-Król and Bauerek, 2015). The markedly higher Fe concentration (5.77 mg l$^{-1}$ in BB and 2.24 mg l$^{-1}$ in BM) is specific to peatlands (between 0.1-6 mg l$^{-1}$ Shotyk, 1988). The dissolved organic carbon (DOC) is constantly elevated, similarly to many peatlands suffering from past drainage. DOC concentrations are higher in the more degraded BB (104-273 mg l$^{-1}$) than in BM (48-133 mg l$^{-1}$).

Based on the field-measured Eh values between 106-297 mV (Table 1), the peatlands are moderately reduced. The lack of redox equilibrium is manifested by the co-occurrence of dissolved oxygen (0.2-3.1 mg l$^{-1}$) and sulfide ions (0.30-2.24 mg l$^{-1}$). Although well below saturation, the presence of O$_2$ indicates much higher redox potential (Eh>900 mV calculated using the pH and T range of the peatland's porewater) than the sulfate-sulfide redox couple (0-60 mV). The deeper peat layers contain more reduced sulfur ions and slightly lower concentrations of sulfates. The metal concentrations are also lower; only As shows higher concentrations in the deeper part of BM (Table S2).

### 3.2. Concentrations of trace elements and sulfur in peat

Four vertical profiles of Zn, Pb, Cd, As, and sulfur distribution are shown in Figs 1 and 2 for BB and BM peatlands, respectively. A metal-enriched layer occurs only a few centimeters below the peat surface. The maximum concentrations (1,038-2,371 mg Zn kg$^{-1}$, 412-1,191 mg Pb kg$^{-1}$, 12-63 mg Cd kg$^{-1}$, and 8.4-23 mg As kg$^{-1}$), the depth of the maxima, and the layer thickness vary between the profiles. As the metal(loid)s deposition history is similar in the two peatlands, all the dissimilarities between metal distribution patterns result from processes occurring within the peat. Firstly, there are variations in the rates of peat accumulation among the profiles. Generally, BM shows higher accumulation rates than BB (Fiałkiewicz-Kozieł et al., 2014). Consequently, the layer appears deeper in BM peatland. Secondly, some downward relocation of the chemical elements is evident from the mutual displacement of the elements concentration maxima and differences in the shape of the concentration curves (Figs 1 and 2). Still, total element loads are comparable (Table S3), indicating limited lateral migration. The higher total loads of Pb and sulfur in BM (Table S3) result from a closer vicinity of the older Pb smelter in Strzybnica (Fig. S1).

A short-range separation of Zn, Cd, and As from Pb is observed in BB. The maximum of Zn, As, and Cd is 2 cm deeper than Pb in BB2 (Fig. 1b). In BB1, Pb forms a broad peak while most Zn, Cd, and As are concentrated in a markedly thinner layer (Fig. 1a). Almost no metal relocation is observed in BM1; only the shape of Zn, Cd, and As differs from the Pb curve (Fig. 2a). An evident mutual separation of Zn, Cd, and As is seen in BM2 (Fig. 2b). The two large Pb peaks in BM profiles probably reflect two deposition events (Fig. 2). The lower peak is within the operation time of Strzybnica Pb smelter (Smieja-Król et al., 2019). Additionally, such a considerable redistribution of Pb in the less disturbed peat is highly unlikely. Pb is regarded as one of the least mobile elements in peat (Shotyk, 1998). The explanation cannot be applied for the lower peaks of Zn, Cd, and As. The additional peaks are exceptionally well developed for As in BM peatland. In addition to the main peaks, two additional sequences are observed, the deepest being below the elevated levels of sulfur concentration (Fig. 2).

**3.3. Sulfides mineralogy and distribution**

SEM analyses document the occurrence of numerous micron to submicron-sized precipitates of ZnS, (Zn,Cd)S, and PbS within the metal-enriched layer in both peatlands (Fig. 3; Figs S4-S6). Their characteristic feature is an intimate association with peat organic matter; the sulfides are attached to or enclosed by peat plant tissues (Fig. 3). ZnS contains a varied amount of Cd (up to 6wt%; see Figs S4 and S5 for Cd distribution in ZnS). Occasionally PbS contains 4-15wt% Cu (Fig. S6). No Fe sulfide was found, nor Fe was detected in Zn and Pb sulfides (Figs. S4-S6).

Spherical aggregates composed of nanometric particles constitute the most basic mode of ZnS precipitation in the peat (Fig. 3a-d). The spheroids' typical size is <1-2 μm, with some up to 4-6 μm. The closer inspection of their surface indicates two varieties of particle arrangement within the aggregates. Some spheroids have a rough, uneven surface (Fig. 3a), resulting from the random arrangement of nanometre-size particles. The second variety is composed of densely packed platelets growing perpendicular to the spheroid's surface, suggesting higher crystallinity and internal ordering of the precipitates. A trigonal arrangement of the platelets is observed (Fig. 3b,c).

Spheroids' overgrowth and colloform infillings of plant cells led to various morphological forms (Fig. 3d-g). Compact pockets of cuboidal shape (Fig. 3f), 20-30 µm in diameter, most probably ZnS mineralized cells of *carex* roots, are most common in the peatlands. The ZnS pockets are irregularly distributed between empty cells along the root; sometimes, only single cells are ZnS impregnated (Fig. 3f). ZnS mineralization in remnants of the Scots pine bark tissue is also identified (Fig. 3g). In that case, whole cells are filled in addition to perfect pseudomorphs formed after Ca oxalate monoclinic crystals (Fig. 3g; Smieja-Król et al. 2014). *Sphagnum* tissues, which are the most abundant macrofossil in the peat (Fiałkiewicz-Kozieł et al., 2014), lack any signs of mineralization.

The authigenic PbS precipitates are much smaller and less frequent. They occur between ZnS spheroids inside plant cells or as flat irregular or square patches in plant cell walls over ZnS (Fig. 3h) or separately (Fig. 3i, Fig. S6).

TEM observations of FIB prepared specimens show the details of two cuboidal root cells extracted from BB1 and BM1 sites, respectively (Fig. S3). Both cells are filled with ZnS exhibiting colloform texture, with a well-developed growth banding (Fig. 4). Growth appears to propagate from the cell's inner highly porous regions outward (Fig. 4) and from the cell wall towards the cell center (Fig. 5a,c). Spheroids, sharing the same texture with the colloform bodies, are also seen (Fig. 6a). Only the BM1 cell contains PbS inclusions (Fig. 7).

An outer rim of markedly lower density (Fig. 5a,c) and higher carbon content (Fig. S7) can be distinguished in the root cells. It has a ribbed texture and a thickness of 0.5-2.0 µm. The HAADF-STEM image shows numerous narrow 4-8 nm, low-density lines with irregular spacing, arranged roughly parallel to the outer cell surface (Fig. 5b). This texture likely formed due to the co-occurrence of ZnS and cellulose microfibrils, a component that provides mechanical strength for plant cells, including roots (Akkerman et al., 2012). A SEAD pattern shows discrete-continuous rings of sphalerite for this part of the cell (Fig. 5a).

The TEM data confirms that the banding in the colloform texture results from differences in density and crystallinity and not from changes in chemical composition (Fig. S7). The low-density bands' thickness is 30-170nm, and the high-density bands are between 40-700nm (Figs 4 and 7a). Radial acicular texture roughly perpendicular to the banding is seen within the high-density bands (Fig. 5c). HRTEM observations reveal large crystalline areas with many planar defects (Fig. 5d) confirmed by a complex SAED pattern (Fig. 5a). Some elongated crystals expand from the denser regions into low-density bands (Figs 4 and 6b). Nano-twins of sphalerite are seen with multiple twin planes oriented parallel to the crystal's elongation (Fig. 6d). The wurtzite structure has not been unambiguously confirmed, although wurtzite-like structural elements are highly possible in the form of closely spaced planar defects. The low-density bands are composed of loosely spaced and randomly oriented rounded crystals. They have 10-30 nm and are generally defect-free (Fig. 6c), similar to the domains observed in the cell wall and the inner low-density regions.

On the contrary, PbS inclusions (Fig. S7) are crystalline and defect-free (Fig. 7a), often with well-developed faces (Fig. 3i, 7a,b). Interestingly, PbS precipitated in the cell wall is segmented by the microfibrils, where the segments differ in orientation (Fig. 7c,d).

### 3.4. Iron speciation

Iron speciation is analyzed in three peat samples per site, covering depth intervals 11-22cm and 11-27cm, at BB1 and BM1 sites, respectively, to explain Fe absence in the authigenic mineralization. Especially that Fe content (1,870-3,450 mg kg$^{-1}$ in BB1 and 1,580-2,020 mg kg$^{-1}$ in BM1 samples) is higher than the content of the trace metals (see section 3.2 for comparison). In both peatlands, the obtained Mössbauer spectra are more complex for the uppermost samples in which four components are distinguished and simpler for the deeper samples showing two quadrupole doublets (Fig. 8). Fe(II) doublets (D4 in Fig. 8) with isomer shift (IS) 0.85 and 0.98 mm s$^{-1}$ and quadrupole splitting (QS) 1.85 mm s$^{-1}$ and 2.04 mm s$^{-1}$ just as Fe(III) doublets (D3) with average IS=0.39 mm s$^{-1}$ and QS=1.17 mm s$^{-1}$ have hyperfine parameters typical for iron in fly ash aluminosilicates from coal-burning in power plants (Ram et al., 1995; Buljan et al. 2015; Smołka-Danielowska et al., 2019). Fly ash spheroidal aluminosilicates were found in the subsurface layers of the peatlands (Smieja-Król et al., 2010; Smieja-Król and Fiałkiewicz-Kozieł, 2015). SEM observations of the Mössbauer analyzed samples confirmed the occurrence of the anthropogenic aluminosilicates in the BB1 and BM1 samples from the 11-12cm depth and a much lower amount in BM1 18-19cm. The particles contain a variable amount of iron, generally within the range of 1-5wt%.

Two quadrupole doublets (D1 and D2) indicating Fe(III) in tetrahedral geometry are seen in all Mössbauer spectra (Fig. 8). Their relative abundance increases in the deeper peat samples. The IS range for D1 is 0.22-0.26 mm s$^{-1}$, and the range of QS is 0.29-0.34 mm s$^{-1}$. The second doublet (D2) shows similar IS values (0.22-0.28 mm s$^{-1}$) and higher QS (0.68-0.98 mm s$^{-1}$). The values are lower than those of known soil minerals (Kodama et al., 1977; Dyar et al., 2006, Murad, 2010, Andrade et al., 2018), including ferrihydrite (IS>0.35 mm s$^{-1}$, QS>0.52 mm s$^{-1}$) or nano-sized goethite (Murad et al., 1988; Van Der Zee et al., 2005; Rancourt et al., 2005), and the low spin Fe(II) in iron sulfides (Mullet et al., 2002; Dyar et al., 2006). The available data for Fe(III) complexes with fulvic acids (IS=0.38 mm s$^{-1}$, QS=0.5-1.0; Kodama et al., 1977) and Fe(III) in organic soils (IS=0.30-0.38, QS=0.58-0.68 mm s$^{-1}$; Mercader et al., 2014; Bhattacharyya et al., 2018) provides parameters which are difficult to distinguish from the inorganic Fe and are distinctly different from that determined in BB and BM peat. On the other hand, laboratory synthesized complex Fe-organic compounds and plant tissues can provide parameters close to that of the peat material, suggesting that the D1 and D2 doublets (Fig. 8) most probably represent organically bound ferric iron (Kovacs et al., 2008; Mishra et al., 2016; Cherkezova-Zheleva et al., 2018).

## 4. Discussion

### 4.1. Controls on the mineral assemblage formation in peat

Reducing conditions are a prerequisite for sulfide mineral formation (Dvorak et al., 1992; Johnson and Hallberg, 2005). The availability of biodegradable organic matter and waterlogged conditions allow sulfate-reducing microorganisms to transform the atmospheric sulfur into reduced species (Bartlett et al., 2009). However, as the peatlands are open systems, they stay only mildly reducing, allowing for constant gas exchange with the atmosphere, i.e., ingress of oxygen (e.g., through plant roots), much enhanced during low water table, and liberation of reduced species ($H_2S$, methane, $CO_2$) (Gorham, 1991; Mitsch and Gosselink, 2007; Lai, 2009). Both dissolved oxygen and sulfide ions were determined in the peat pore water (Table 1 and Table S2). We suggest that the formation of physicochemical micro-gradients drives the subsurface sulfide mineralization. A diffusion-limited solute exchange can occur within the complex network of peat pores (Kirk, 2004; Rezanezhad et al., 2016) and heterogeneous distribution of biodegradable organic compounds. While clustering in microsites, sulfate reducers locally build up sulfide concentration leading to mineral precipitation in the surrounding. The mobile metal and sulfate ions diffuse to the microsites where the colonies of sulfate reducers develop. The SEM observations confirm that the mineralization is plant tissue-dependent. The degradation-resistant *sphagnum* moss remains unmineralized (Smieja-Król et al., 2010; Hájek et al., 2011), while Zn sulfides commonly infill *carex* roots and Scots pine bark tissues (Fig. 3f,g). Although a direct association of the precipitates with microorganisms (e.g., mineralized bacteria cells or microbial biofilms) was not observed in the peat, the clustering of the metal sulfides within neighboring plant cells (Fig. 3d-g; Figs S4 and S5) supports the concept of metal sulfide precipitation close to (mm to µm range) a source of sulfide ions.

The organic matter's complex chemical structure (in both dissolved- and solid-state) enables several mechanisms to be involved in metals behavior and mineral reaction kinetics of both dissolution and precipitation, i.e., ion-exchange, complexation with ligands, the formation of out- and inner-sphere complexes, and chemisorption (Brown et al., 2000). We postulate that the preferential precipitation of ZnS and (Zn,Cd)S over other sulfides results directly from the complex metal-organic interactions. Although galena precipitation was confirmed during this study (Figs 3h,i and Fig.7), the phase is much less common than sphalerite despite comparable Zn and Pb content in the peatlands (Table S3). Additionally, while Fe content is higher than that of Pb and Zn and uniformly distributed (compare Smieja-Król et al., 2010), secondary authigenic Fe minerals are utterly absent in the peat. Copper occurring in the peat in a low amount (up to 39 mg kg$^{-1}$; see Smieja-Król et al., 2019, for complete Cu profiles) tends to be associated with galena (Fig. S6).

The sporadic occurrence of galena being in contrast with the high Pb content (up to 1,191 mg Pb kg$^{-1}$) can be explained by the high affinity of Pb to organic matter. Pb is considered the least-mobile element in peat, retained through physiochemical binding to organic matter (Vile et al.,1999; Brown et al., 2000; de Vleeschouwer et al., 2009, Lourie and Gjengedal, 2011). Also, the dissolved Pb is effectively complexed by DOC under a range of pH conditions (Lazerte et al., 1989; Rothwell et al., 2011), limiting the availability of free Pb ions to react with sulfide. The extent and mechanism of Cu, Zn, and Cd immobilization after atmospheric deposition is less clear (e.g., de Vleeschouwer et al., 2009), and several competitive processes are considered, including plant uptake (Shotyk et al., 2019). Experimental studies show that the binding capacities of peat for the metals follow the order $Cu^{2+} > Zn^{2+} > Cd^{2+}$ (Twardowska and Kyzioł, 1996). Only ~30% of Zn ions are sorbed on *Sphagnum* peat from solution in monometallic batch experiments (Twardowska et al., 1999). Likewise, more Zn and Cd are in a free ion form in DOC-rich pore waters than Pb and Cu (Lazerte et al., 1989; Linton et al., 2007), favoring the precipitation of Zn-Cd sulfides over Pb. The different immobilization mechanisms are seen in the metal concentration profiles (Figs 1 and 2), where Zn and Cd are separated from Pb. Cu precipitation or incorporation in sulfides is primarily limited by low Cu content in the peat.

The absence of iron in sulfides observed using SEM (Figs S4 and S5) stays in agreement with Mossbauer spectroscopy results (Fig. 8), indicating that organically bound ferric ($Fe^{3+}$) iron is the main form of Fe at the depth range of the authigenic mineralization in the studied peatlands. The finding is supported by numerous studies showing that organic matter can stabilize the ferric form (Karlsson et al., 2008; Prietzel et al., 2009; Beckler et al., 2015; Kügler 2019), allowing for its occurrence under reducing conditions, much below the redox stability of Fe(III) (Steinmann, Shotyk, 1995; Bhattacharyya et al., 2018). Previous studies show that Fe is mainly delivered through dust deposition to the peatlands (Smieja-Król et al., 2010). After the dissolution of the dust particles (Smieja-Król et al., 2010), the organic matter preserves the ferric state of Fe and suppresses the Fe(II) ions activity. This then favors the precipitation of the less common sulfides.

However, authigenic pyrite commonly occurs in minerotrophic peatlands fed by groundwaters (e.g., Papunen 1966; Chague-Goff et al., 1996; Rydelek 2013), and iron was found to be incorporated in Zn sulfides (Sonke et al., 2002; Awid-Pascual et al., 2015). That can be explained by high renewal rates of anoxic and iron (II) rich groundwaters, which oversaturate the capacity of DOC to form complexes with Fe, leading eventually to Fe sulfides precipitation. On the contrary, the studied peatlands are rich in dissolved organic matter and fed mainly by precipitation with limited groundwater inflow and outflow (Smieja-Król and Bauerek, 2015).

Arsenic relocation into deeper peat layers is more pronounced than the metals, assuming a common deposition history of the pollutants. The downward As splitting, especially evident in the wetter BM peatland, closely resembles Liesegang bands formation (Fig. 2), i.e., diffusion in porous media and periodic immobilization through precipitation, absorption, and/or transformation to less mobile form (Jamtveit and Hammer, 2012). Arsenic was not detected in the sulfides (see Figs S4 and S5) despite comparable to Cd contents in peat (Table S3) and higher porewater concentrations (Table 1 and Table S2). The work of Drahota et al. (2017) suggests a requirement of more reducing conditions for the reduction of As(V) to As (III) in excess of organic matter. Organo-arsenic compounds are known to form in peatlands subjected to gross air pollution (Mikutta and Rothwell, 2016).

### 4.2. Model of crystallization and coarsening of the observed sulfides

ZnS is extremely fine-grained when precipitated at ambient conditions. Both laboratory and field studies indicate ZnS particulates dimensions within the range of 2-12 nm (e.g., Huang et al., 2003; Moreau et al., 2004; Castillo et al., 2012; Bodo et al., 2012; Xu et al., 2016).

Organically capped ZnS is slightly coarser than ZnS obtained in purely inorganic conditions (Huang et al., 2003; Xu et al., 2016). The difficulty in crystallization is probably related to the surface instability of the ZnS growth units resulting from the unit's dipole-dipole interactions with surrounding water molecules (Xu et al., 2016).

In this study, the colloform and spherical aggregates are composed of nano-particulates and submicrometre defected sphalerite crystals (Fig. 6 c,d). Two-step ZnS growth is postulated. The first step leads to the formation of 10-30nm particles, observed in the lower density regions by TEM (Fig. 6c) and as rough, uneven surface of spheroids in SEM images (Fig. 3a). They are larger than particles synthesized in the laboratory but comparable to the dimensions of sphalerite (48 nm) documented by Yoon et al. (2012) in minerotrophic peatland using synchrotron-based XRD. The larger dimensions of the naturally forming crystallites result probably from slow crystallization in low supersaturation conditions stabilized by organic molecules.

A later step of crystal growth involving oriented attachment mechanism explains the occurrence of large, highly defected crystalline areas, forming platelets as observed by SEM (compare Fig. 3b,c and Fig. 5c,d). The degree of crystallinity resembles hydrothermally coarsened ZnS in laboratory conditions (Huang et al., 2003; Huang and Banfield, 2005). Although the exact time of reactants deposition and mineral precipitation is unknown, considering the much-reduced smelter emission since 1990 (Fig. S2), the precipitates are probably older than 30 years. Consequently, time alone might be responsible for the improved crystallinity, as the temperature does not exceed 25°C (Table 1). The time-dependent decay of organic molecules probably helps in releasing the initially capped ZnS nanoparticles so that they can rearrange into larger crystals. The exception is the degradation-resistant cellulose, preventing nanoparticles reorganization in the outermost cell layer. The enhanced crystallinity and aggregate coarsening at the early stage make the sphalerite more resistant to oxidative dissolution.

Although galena is the easiest sulfide to make by bacterial sulfate reduction, according to Baas Becking and Moore (1961), and is commonly biosynthesized in a laboratory (Öcal et al., 2020; Staicu et al., 2020), its precipitation in nature is rarely observed. Only a few occurrences are confirmed (Sonke et al., 2002; Awid-Pascual et al., 2015; Smieja-Król et al., 2015). The present study shows that the growth mechanism of galena differs from that of sphalerite despite common precipitation conditions. Galena forms individual defect-free crystals much larger than that of sphalerite (Fig. 7a). They resemble crystals grown in porous media by slow diffusion of reactants approaching equilibrium concentration (García-Ruíz, 1986). As discussed previously (section 4.1.), the Pb ion availability is reduced by complexation with organic molecules.

## 5. Implications

Terrestrial, near-surface, organic-rich environments are conducive to the precipitation and preservation of Zn-Cd-Pb sulfides. In the study, a coherent near-surface authigenic ZnS, (Zn,Cd)S, and PbS mineralization was shown to extend over the area encompassed by two peatlands as a consequence of air deposited pollution transformation into secondary reduced phases.

The depositional environment described here is remarkably similar to metal accumulations known from geological records in terms of mineral assemblage, chemical composition, petrographic features, and the overall deposit geometry (e.g., Craig and Vaughan, 1994; Large et al., 1998; Barrie et al., 2009; Kucha et al., 2010; Leach et al., 2010; Wen et al., 2016). The spherical and colloform aggregates of sphalerite contain variable amounts of Cd (Fig. S5) and accommodate their outer shape to the available space, i.e., pseudomorphs are common (Fig 3e-g). The polycrystalline sphalerite is associated with much larger well-shaped galena crystals (Figs 3i and 7a). The mineralization is sphalerite-rich relative to galena and is stratiform. After diagenesis, compaction, and peat degradation, the early sulfidic mineralization might survive as a thin high-grade Zn-Cd-Pb sulfide layer in sandstones.

The implications of our observations are twofold. First, we show that industrial atmospheric emissions are re-sequestered into secondary sulfides in addition to the well-documented metal complexation onto organic matter and retainment in resistant primary particles (Twardowska and Kyzioł, 1996; Vile et al.,1999; Brown et al., 2000; Rausch et al., 2005). The authigenic mineralization, hidden in plant remains, has increased resistance over time. The Cd-containing ZnS aggregates have a complex polycrystalline sphalerite structure much more advanced than typically obtained during low-temperature synthesis or observed in other modern occurrences. Platy highly-disordered radially-aggregated submicrometre crystals (Figs 3b,c, and 6d) develop within the time constraints of several decades in the cold (~15°C) and acid (pH 3.4-4.4) peat. The galena well- developed crystallinity (Figs 3i and 7a) increases its stability against oxidation. Consequently, peatlands may serve as sinks of contaminants, retarding the cycling of metals. The human-driven enrichments of industrially important metals have dramatically increased in the twentieth century, giving a global signal in geological archives (Nriagu and Pacyna, 1988;

Rauch and Pacyna, 2009; Thorne et al., 2018). When the mineralization is high-grade and substantial in volume, the polluted sites can be considered as human-induced ore deposits, adding to the list of significant anthropogenic impacts on Earth's geology, leading recently to the recognition and formalization of the Anthropocene epoch (Waters et al., 2016; Waters et al., 2018).

Secondly, the polluted sites can be used to understand and reconstruct the genesis and the formation mechanisms of ancient deposits. Our findings highlight the role organic matter exerts on the mineralization process. Reagents diffusing through porous peat, selective formation of metal-organic complexes, and heterogeneous susceptibility to biodegradation are the major impactors of the mineral assemblage formation. Different affinity to organic matter of the dominating metals (Zn, Pb, and Fe) influences the mineral assemblage. Contrariwise to the lower solubility and high content in the peat (Table S3), PbS is much less frequent than ZnS. Fe is excluded from the early mineralization stage due to its preservation in the ferric oxidation state by organic matter.

It is inferred from the study that the potential environments for nonferrous metal sulfide accumulation in recent and past geological settings can be much wider than formerly realized. The mineralization might form and sustain in the near-surface environment. No strictly anaerobic conditions are required. The ingress of oxygen through precipitation, plant root system, or water table drawdown is not detrimental to the metal sulfides. An earlier study has shown that even a very thin layer (<30 cm) of biodegradable organic matter is sufficient for metal sulfide crystallization and persistence (Smieja-Król et al., 2015). Potentially convenient are lacustrine, alluvial, and coastal plains, as well as fluvial-delta depositional systems. All the environments are suitable for peat accumulation. Additionally, shallow marine environments rich in readily biodegradable organic matter provide similar conditions. In past geological settings, the mineral-forming reactants might be delivered with surface waters carrying material from the bedrock, ancient deposits weathering, and/or origin from volcanic emissions. The last source is known to inject pointwise vast amounts of sulfur and trace elements into the atmosphere, especially in the pre-anthropogenic times (Hinkley 1991; Hinkley et al., 1999; Masotta et al., 2016; Edmonds and Mather, 2017).

Acknowledgments

This work was supported by the National Science Centre, Poland [grant number 2016/23/B/ST10/00781]. Barbara Liszka, Marek Król, and Krzysztof Kozieł are thanked for their help in peat sampling. The authors thank the editor and reviewers for their valuable comments on the first version of this paper.

Table 1. Mean values (ranges in brackets) of physical and chemical properties of shallow (5-40 cm) peat porewater during water-logged conditions in BB and BM peatlands.

|  | BB1 | BB2 | BM1 | BM2 |
|---|---|---|---|---|
| WTD*, $cm$ | -2.2 (-7-0) | -2.3 (-7-0) | 18 (2-30) | 15 (-2-26) |
| T, $°C$ | 10 (4.4-18) | 11 (4.5-18) | 15 (7.2-25) | 15 (8.2-24) |
| pH | 3.7 (3.5-4.0) | 3.7 (3.5-3.9) | 4.0 (3.7-4.4) | 4.0 (3.7-4.3) |
| EC, $\mu S\ cm^{-1}$ | 99.5 (92-101) | 95 (80-134) | 55 (34-86) | 56 (39-64) |
| Eh, $mV$ | 258 (255-260) | 266 (254-283) | 236 (179-294) | 244 (185-297) |
| $O_2$, $mg\ l^{-1}$ | 1.8 (1.7-2.0) | 2.7 (1.6-4.0) | 1.2 (0.2-3.1) | 1.0 (0.3-1.9) |
| DOC, $mg\ l^{-1}$ | 147 (113-207) | 117 (85-172) | 57 (48-73) | 63 (60-98) |
| sulphate, $mg\ l^{-1}$ | 10.9 (5.0-25) | 9.3 (3.7-23) | 7.4 (1.6-23) | 7.0 (1.5-21) |
| sulfide, $mg\ l^{-1}$ | 0.81 (0.28-1.6) | 0.93 (0.4-2.2) | 0.43 (0.27-0.80) | 0.36 (0.24-0.56) |
| chloride, $mg\ l^{-1}$ | 4.8 (4.0-6.2) | 4.8 (2.9-7.1) | 3.6 (1.0-5.8) | 3.7 (1.8-5.9) |
| Fe, $mg\ l^{-1}$ | 6.50 (5.20-6.91) | 5.03 (4.01-6.85) | 2.83 (1.63-4.26) | 2.69 (1.43-4.46) |
| Zn, $mg\ l^{-1}$ | 0.56 (0.38-0.72) | 0.63 (0.46-0.88) | 0.21 (0.13-0.30) | 0.16 (0.11-0.25) |
| Pb, $\mu g\ l^{-1}$ | 45.0 (22.8-109) | 45.0 (14.3-128) | 6.27 (2.25-11.5) | 9.57 (5.16-16.4) |
| Cd, $\mu g\ l^{-1}$ | 2.30 (1.03-5.54) | 2.24 (1.06-6.19) | 0.69 (0.41-1.40) | 0.57 (0.31-1.23) |
| As, $\mu g\ l^{-1}$ | 13.3 (9.51-19.4) | 11.0 (7.14-16.8) | 4.92 (2.31-6.09) | 5.51 (3.12-7.97) |

*WTD – water table depth relative to peat surface

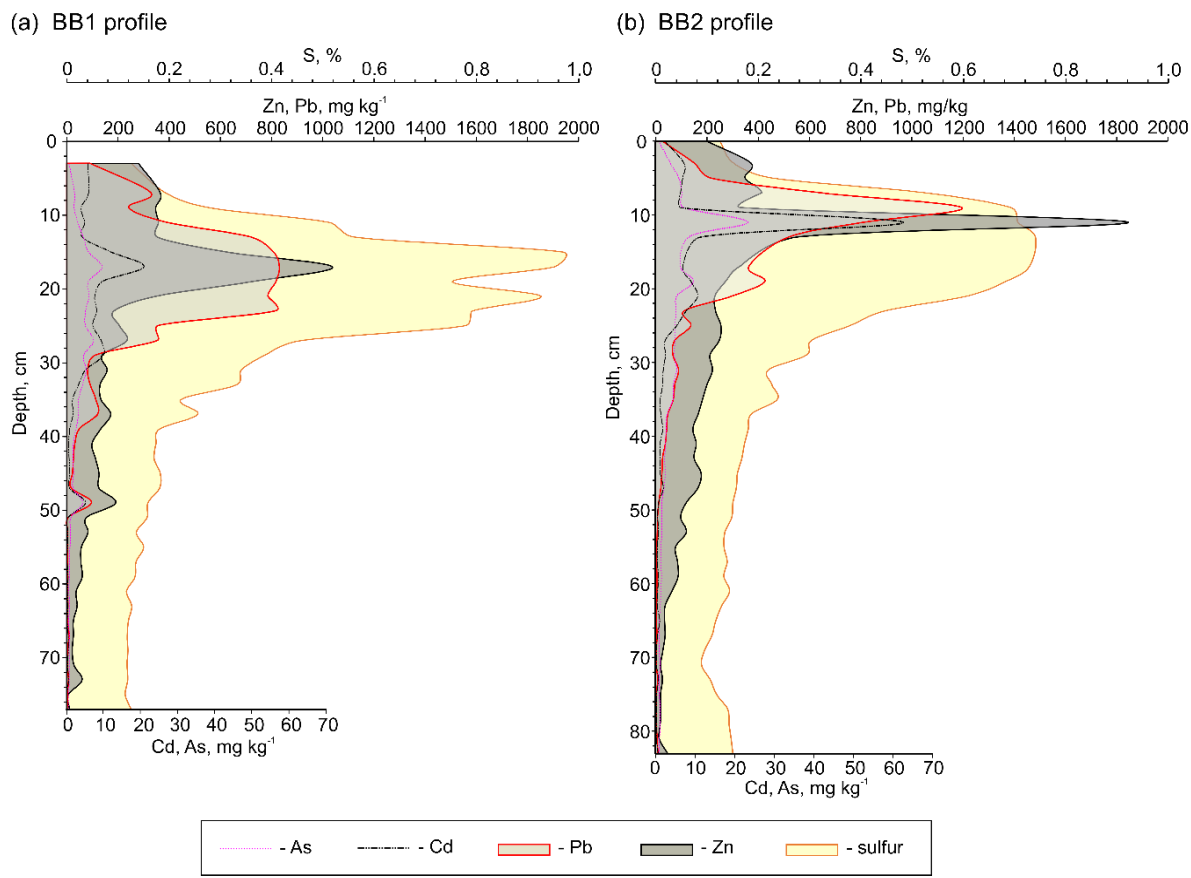

Fig. 1. Pb, Zn, Cd, As, and S concentrations in (a) BB1 and (b) BB2 profiles.

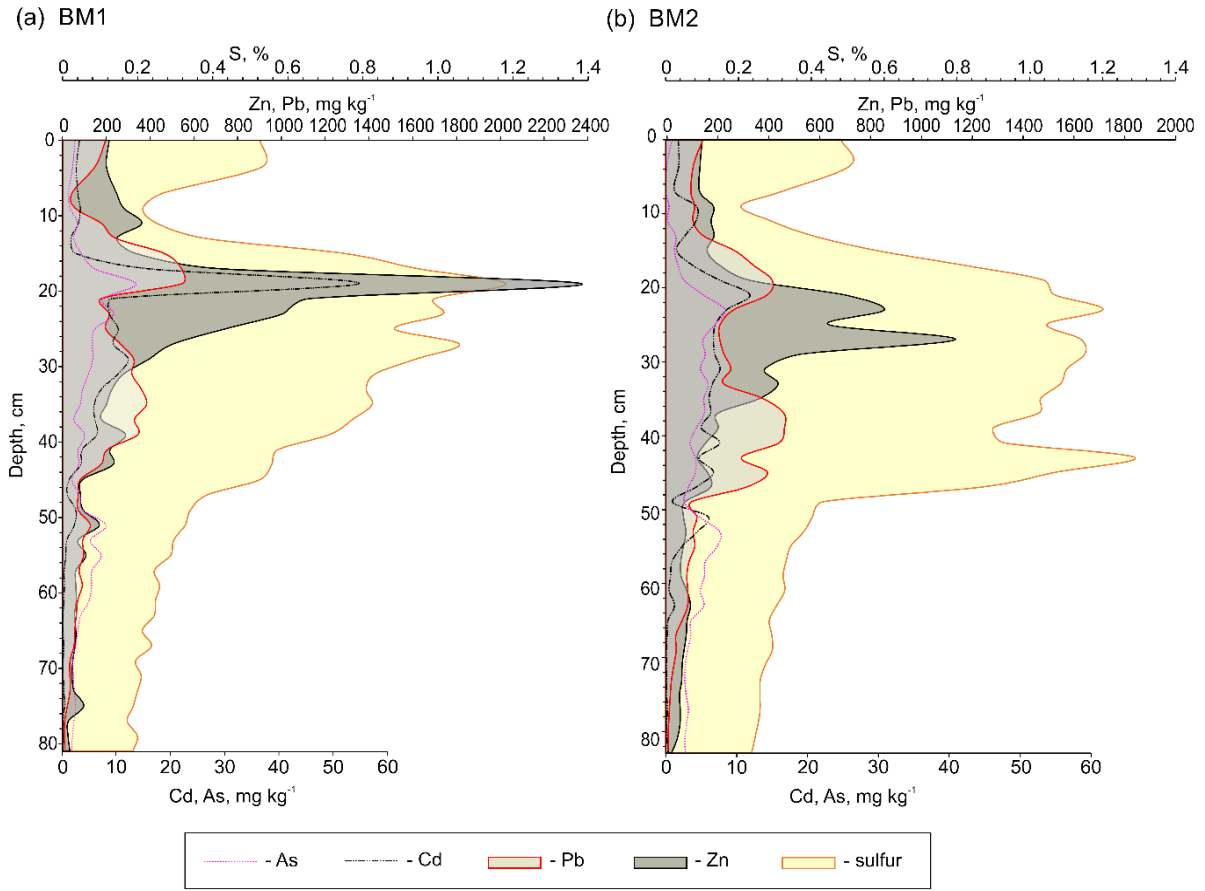

Fig. 2. Pb, Zn, Cd, As, and S concentrations in (a) BM1 and (b) BM2 profiles.

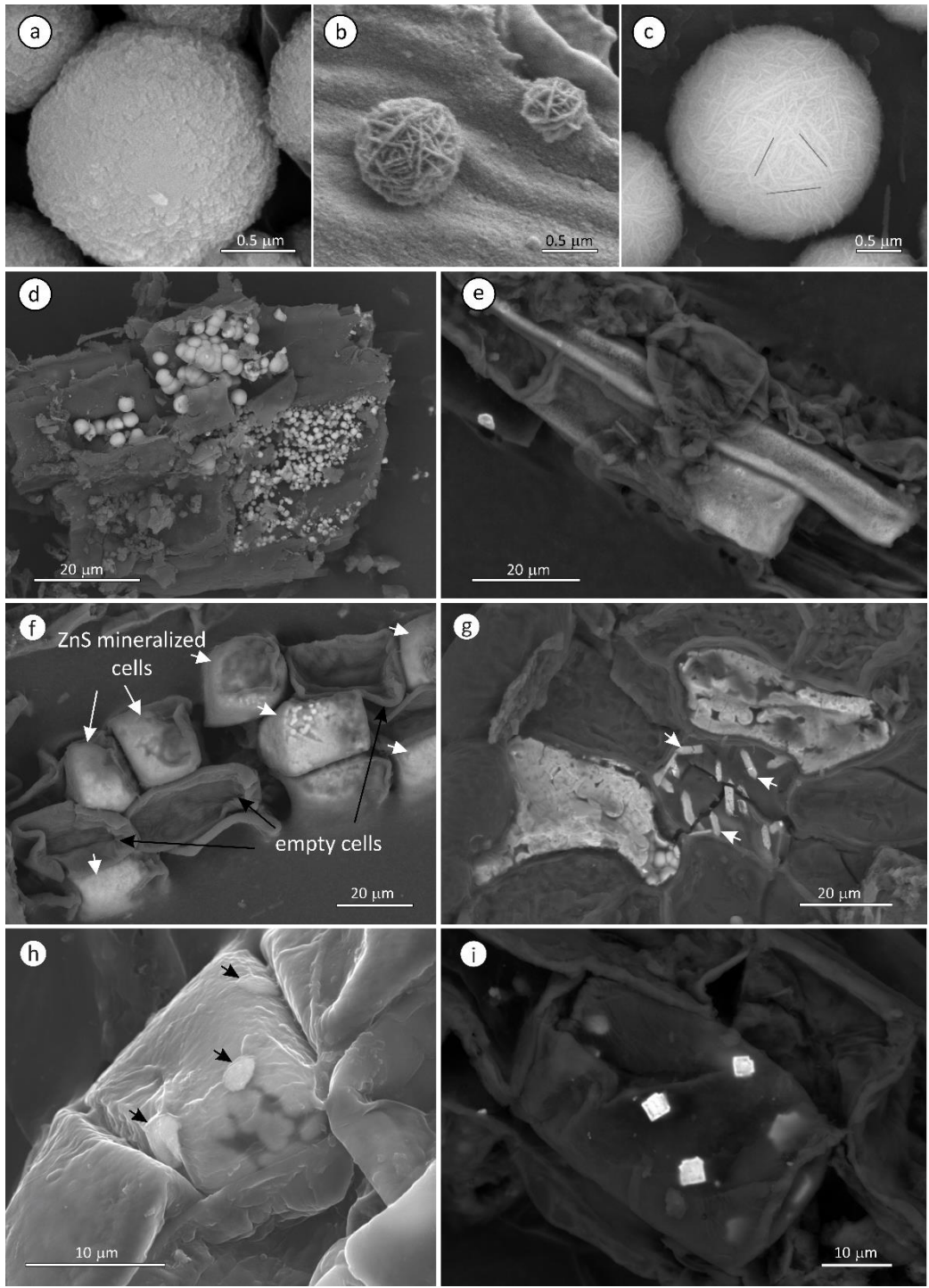

Fig. 3. SEM images of authigenic ZnS and PbS precipitates. (a) ZnS spheroid with rough surface (BSE; BB1, 19-20 cm); (b) rosette-like ZnS spheroid (SE; BB1, 15-16cm); (c) large ZnS spheroid with trigonal platelets arrangement (BSE; BB1, 19-20 cm); (d) ZnS spheroids grouped in a loose aggregate (BSE; BM1 27-28cm); (e) ZnS mineralized plant cells (BSE; BB2, 11-12cm); (f) a root fragment with randomly distributed ZnS-filled cells (BSE; BM1, 19-20 cm); (g) ZnS-mineralized bark cells of scots pine, ZnS pseudomorphs after Ca oxalate are indicated by arrows (BSE; BB1, 17-18 cm); (h) PbS within a cell wall (arrows) of ZnS-mineralized cell (SE; BM1 17-18 cm); (i) square PbS patches inside a plant cell (BSE; BM1 20-22cm).

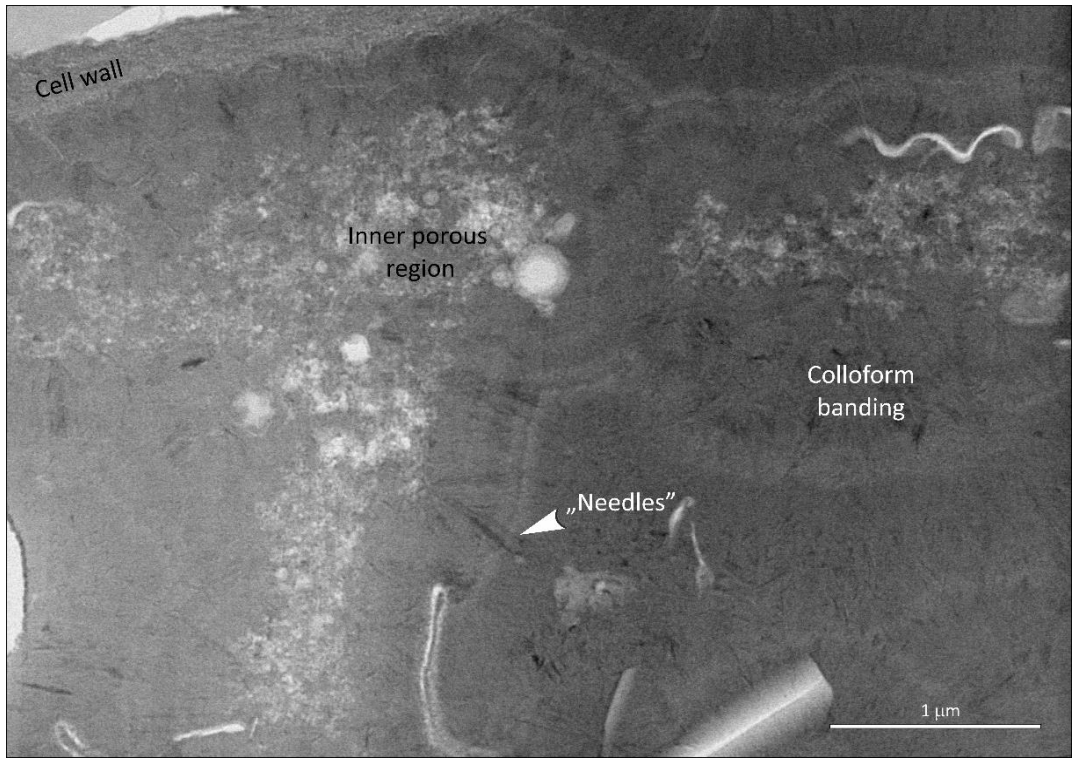

Fig. 4. STEM-BF wide-field view of main textural varieties of ZnS mineralization in root cell. Needle-like crystalline areas are seen in the colloform bands (arrow) due to stronger diffraction contrast (FIB prepared thin section; BB1, 20-21 cm).

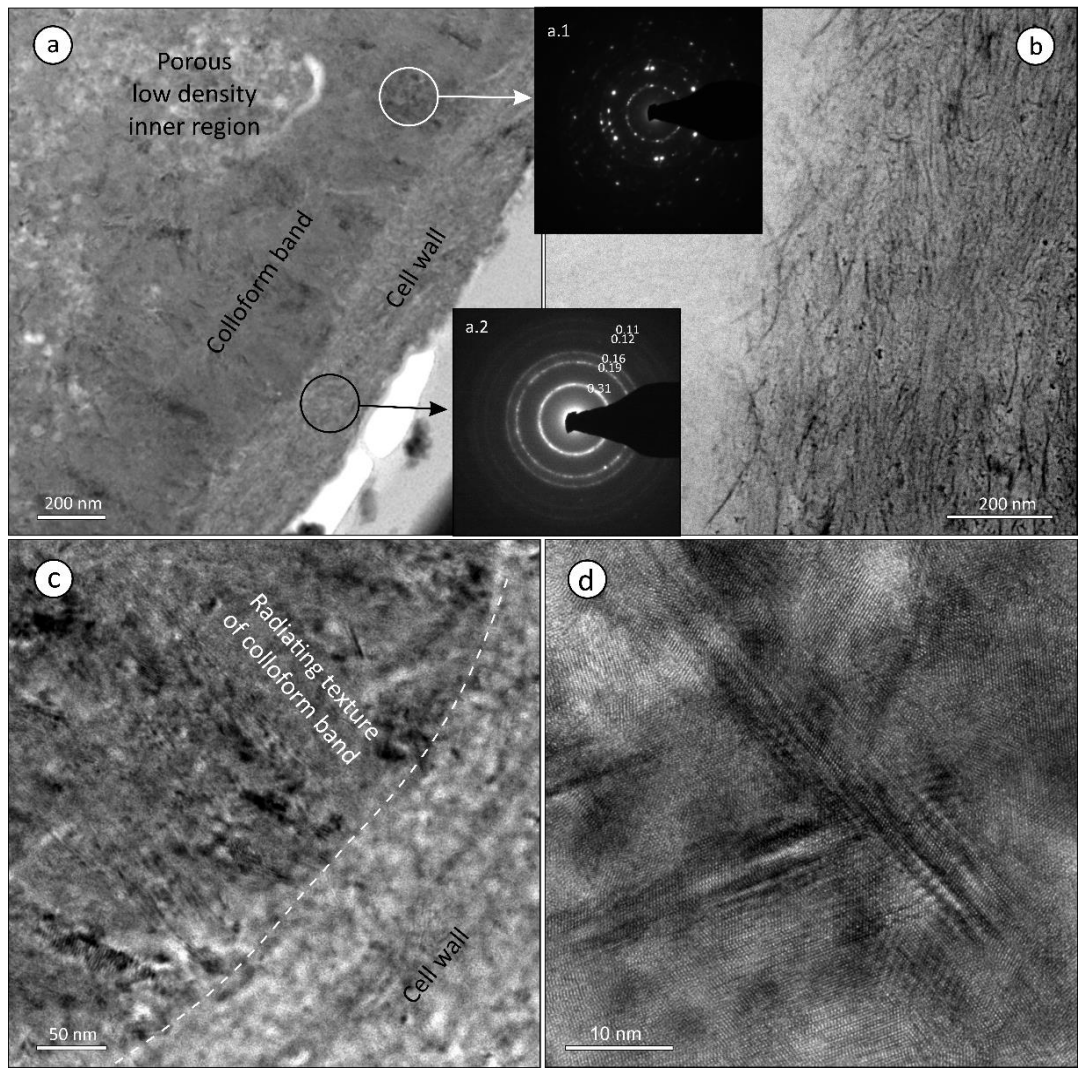

Fig. 5. TEM images of ZnS mineralized root cell (FIB prepared thin section; BB1, 20-21 cm). (a) The cell wall (right), colloform high-density band (middle), and the inner porous region (left) showing differences in density and crystallinity (TEM-BF); (a.1, a.2) the corresponding electron diffraction patterns (aperture diameter 170 nm) of nano-crystalline sphalerite; (b) STEM-HAADF image of the boundary between fibrous cell wall (right) and the uniformly dense colloform band (left); (c) close-up of (a), TEM-BF image underlining the radiating texture of the colloform variety; (d) HRTEM image of highly-defected sphalerite crystal.

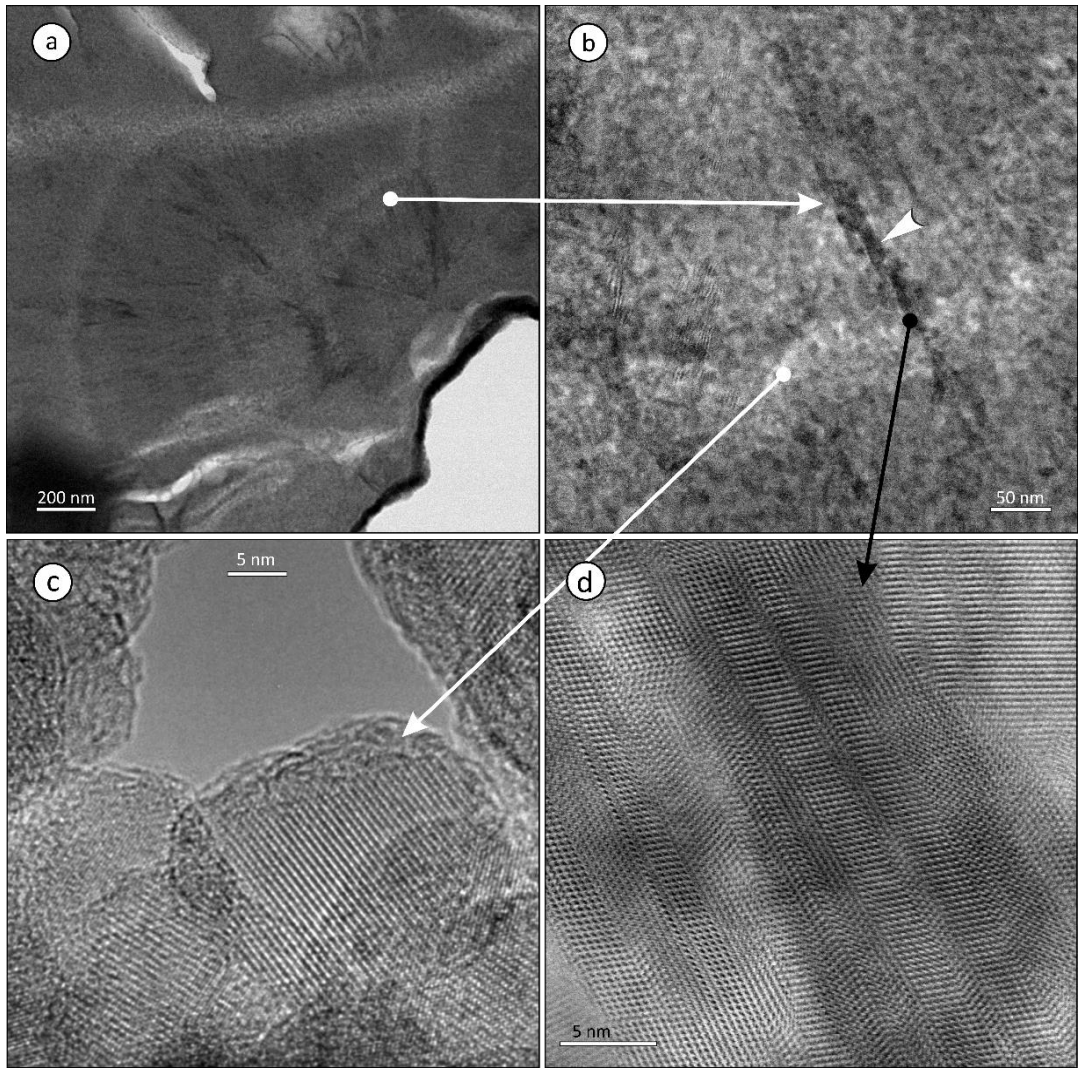

Fig. 6. TEM images of ZnS spheroid (FIB thin section, BB1, 20-21 cm). (a) STEM-BF image of a banded spheroid aggregate (a low-density linear structure formed after some organic filament is seen in the upper part of the image); (b) STEM-BF image of alternating low-, high-density bands showing a crystal (arrow) propagating through the bands; (c) sphalerite crystals from the low-density band (HRTEM); the d-spacing for the revealed face is 0.31 nm, consistent with sphalerite {111} planes; (d) large sphalerite crystal with multiple parallel twin planes and stacking faults (STEM-BF).

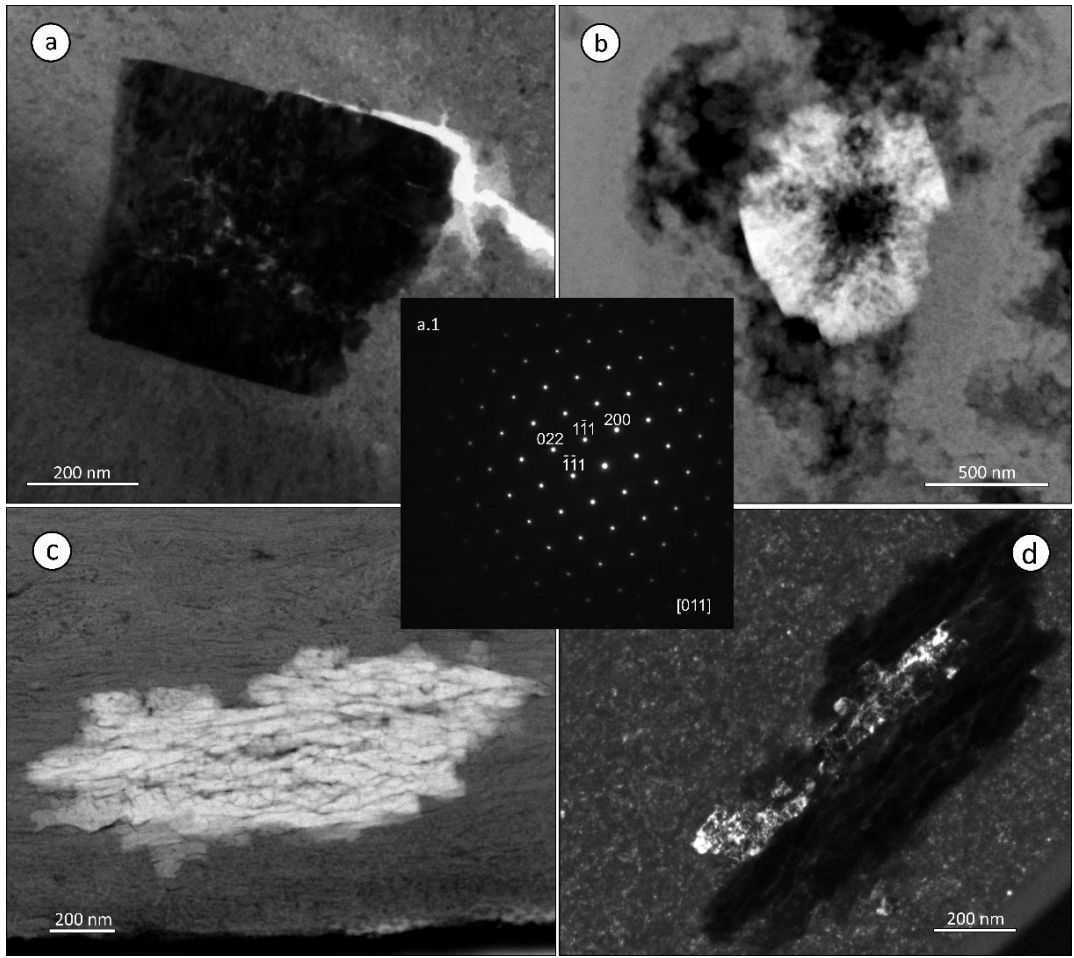

Fig. 7. TEM images of galena inclusions in ZnS mineralized root cell (BM1, 18-19 cm). (a) Euhedral galena crystal (TEM-BF) and (a.1) the corresponding SAED pattern; (b) skeletal crystal (light) located in the inner porous part of the cell (STEM-HAADF); (c) galena grain (light) inside ZnS mineralized cell wall, STEM-HAADF image; (d) TEM-DF view of galena particle from (c) showing domains differing in orientation; small (10-30 nm) sphalerite crystals are forming the image background.

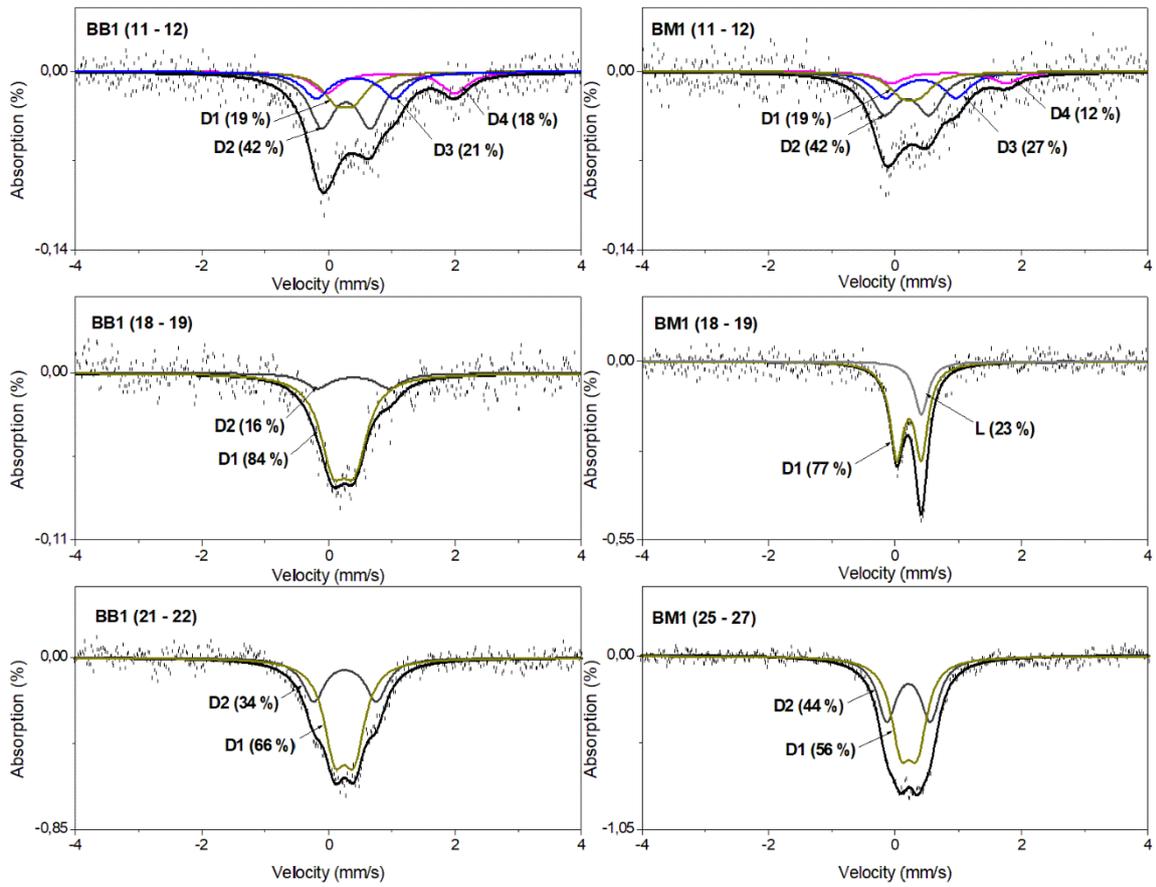

Fig. 8. Mössbauer spectra of peat samples from BB1 and BM1 sites. Black points denote experimental data. The fitted subspectra (colored lines), their assignment (D - doublet, L - singlet), and contributions are presented together with the overall fit (black line).

Table S1. Quality control for the metal(loid)s recovery using NIMT/UOE/FM/001 reference material (Yafa et al., 2004) and sulfur measurements using Eltra Coal Standard (92510-27). Laboratory-provided information on measurement precision (relative standard deviation of reproducibility – $RSD_R$) and analyte recovery are included.

| NIMT/UOE/FM/001 | Certified | Measured (n=4) | $RSD_R$ | NIST 1643-e recovery |
|---|---|---|---|---|
| Element | (mg kg$^{-1}$) | | % | % |
| As | 2.44±0.55 | 3.02±0.5 | 5 | 87.99 |
| Cd | 0.38±0.08 | 0.28±0.04 | 5.3 | 96.81 |
| Pb | 174±8 | 138.9±7 | 13.9 | 105.67 |
| Zn | 28.6±1.9 | 27.36±2.5 | 11.4 | 80.27 |
| Fe | 921±84 | 824±51 | 5.4 | 92.21* |
| Eltra coal standard | Certified | Measured (n=5) | | |
| Element | % | | | |
| S | 0.73±0.02 | 0.734±0.009 | | |

*SPS-SW2 Batch 120 was used for Fe recovery determination.

Table S2. Mean values (ranges in brackets) of physical and chemical properties of peat porewater from deeper peat layer (60-100 cm) during water-logged conditions in BB and BM peatlands.

|  | BB1 | BB2 | BM1 | BM2 |
|---|---|---|---|---|
| T, °C | 9.8 (3.8-17) | 10 (4.2-16) | 11.4 (7.1-17) | 12 (7.5-18) |
| pH | 3.4 (3.4-3.6) | 3.5 (3.4-3.8) | 4.0 (3.7-4.4) | 4.2 (3.8-4.5) |
| EC, $\mu S\ cm^{-1}$ | 129 (80-142) | 111 (87-121) | 80 (71-141) | 78 (71-105) |
| Eh, $mV$ | 236 (189-262) | 253 (231-268) | 190 (148-205) | 145 (106-180) |
| $O_2$, $mg\ l^{-1}$ | 1.2 (0.8-1.8) | 1.9 (1.3-2.3) | 1.1 (0.2-2.1) | 0.7 (0.3-2.0) |
| DOC, $mg\ l^{-1}$ | 212 (114-282) | 163 (104-273) | 92 (50-127) | 118 (81-133) |
| sulphate, $mg\ l^{-1}$ | 7.6 (4.1-20) | 8.1 (3.2-21) | 7.6 (3.1-23) | 8.0 (2.4-20) |
| sulfide, $mg\ l^{-1}$ | 1.21 (0.68-1.80) | 1.00 (0.64-1.60) | 0.52 (0.40-0.80) | 0.68 (0.52-0.80) |
| chloride, $mg\ l^{-1}$ | 6.4 (5.1-8.2) | 5.1 (4.4-6.2) | 5.9 (3.9-9.1) | 6.0 (4.3-9.2) |
| Fe, $mg\ l^{-1}$ | 2.72 (1.83-3.46) | 3.07 (2.83-3.42) | 2.61 (1.86-3.46) | 3.30 (2.44-6.47) |
| Zn, $mg\ l^{-1}$ | 0.14 (0.11-0.20) | 0.22 (0.10-0.38) | 0.19 (0.12-0.23) | 0.12 (0.06-0.20) |
| Pb, $\mu g\ l^{-1}$ | 7.38 (3.04-9.87) | 8.18 (3.48-18.7) | 9.08 (2.60-18.5) | 7.86 (5.59-11.9) |
| Cd, $\mu g\ l^{-1}$ | 0.62 (0.60-0.73) | 0.77 (0.41-1.11) | 0.43 (0.26-0.77) | 0.38 (0.21-0.58) |
| As, $\mu g\ l^{-1}$ | 10.1 (8.42-11.6) | 9.66 (7.60-12.5) | 5.38 (2.51-7.31) | 7.96 (4.08-12.2) |

Table S3. Total cumulative pollution loads and the mean peat bulk density of BB and BM cores estimated in the ~1m upper peat layer using formulas provided by Miszczak et al. (2020). Peat bulk density values were taken from Fiałkiewicz-Kozieł et al. (2014).

| Core | Zn | Pb | Cd | As | S | Bulk density |
|---|---|---|---|---|---|---|
| | \multicolumn{5}{c}{g m$^{-3}$} | g cm$^{-3}$ |
| BB1 | 21.5 | 18.7 | 0.39 | 0.37 | 362 | 0.14 |
| BB2 | 25.6 | 14.3 | 0.42 | 0.45 | 367 | 0.17 |
| BM1 | 31.6 | 19.3 | 0.50 | 0.54 | 592 | 0.14 |
| BM2 | 23.4 | 19.3 | 0.39 | 0.51 | 719 | 0.14 |

Calculation of total cumulative pollution loads (after Miszczak et al., 2020):

1. The calculation of incremental load of the element in a layer $n$ ($L_n$):

$$L_n \text{ (g m}^{-2}\text{)} = \text{layer bulk density (g cm}^{-3}\text{)} * \text{layer thickness (cm)} * \text{element concentration (mg kg}^{-1}\text{) in the layer} * 10^{-2}$$

2. Total retained element load TL (g m$^{-3}$) in peat profile:

$$TL \text{ (g m}^{-3}\text{)} = \Sigma L_n \text{ (g m}^{-2}\text{)}/\text{profile thickness (m)}.$$

A mean value of the concentration from adjacent layers was assigned to the unmeasured samples.

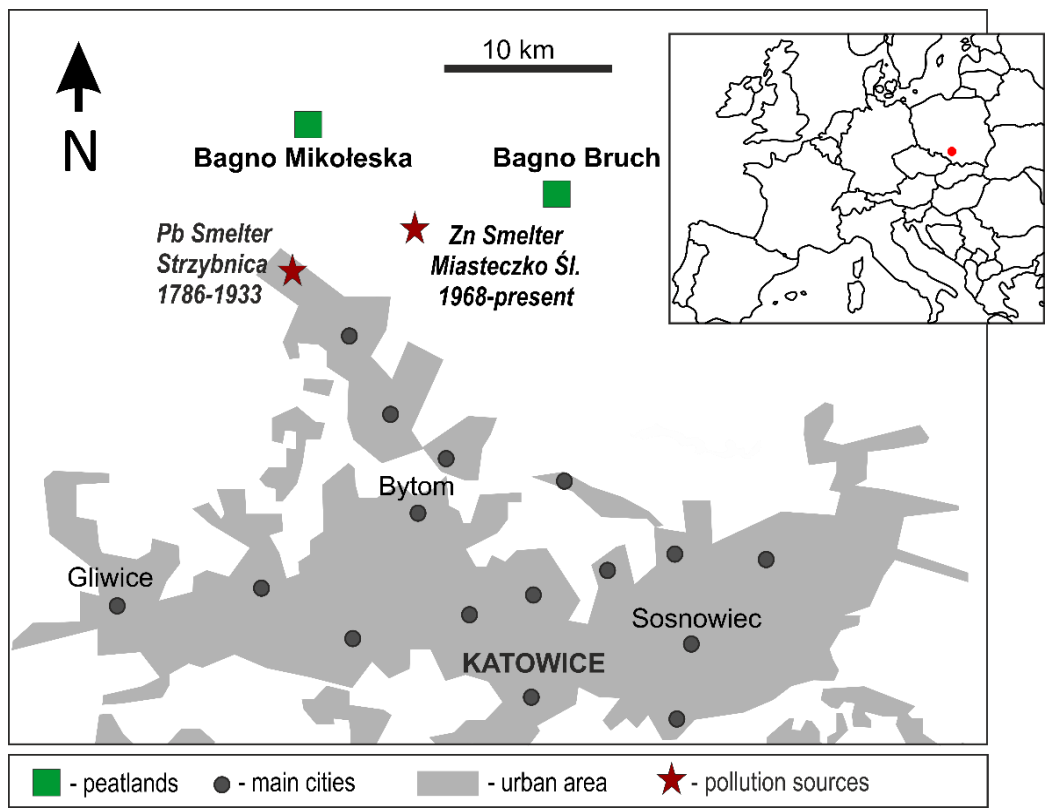

Fig. S1. The location of Bagno Bruch (BB) and Bagno Mikołeska (BM) peatlands in relation to Pb and Zn smelters and the Upper Silesia urban conurbation.

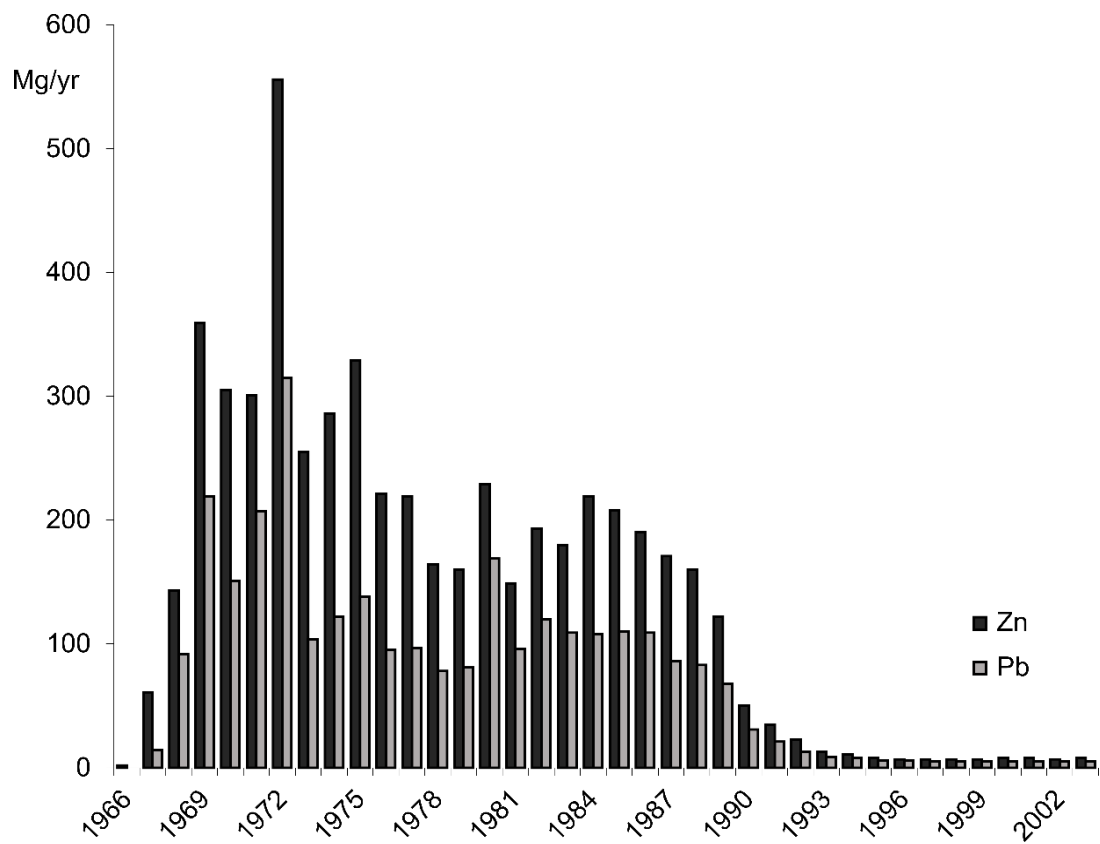

Fig. S2. Yearly zinc and lead emission from the Zinc Smelter Miasteczko Śląskie in 1966-2003 (based on data from Gerold-Śmietańska, 2007).

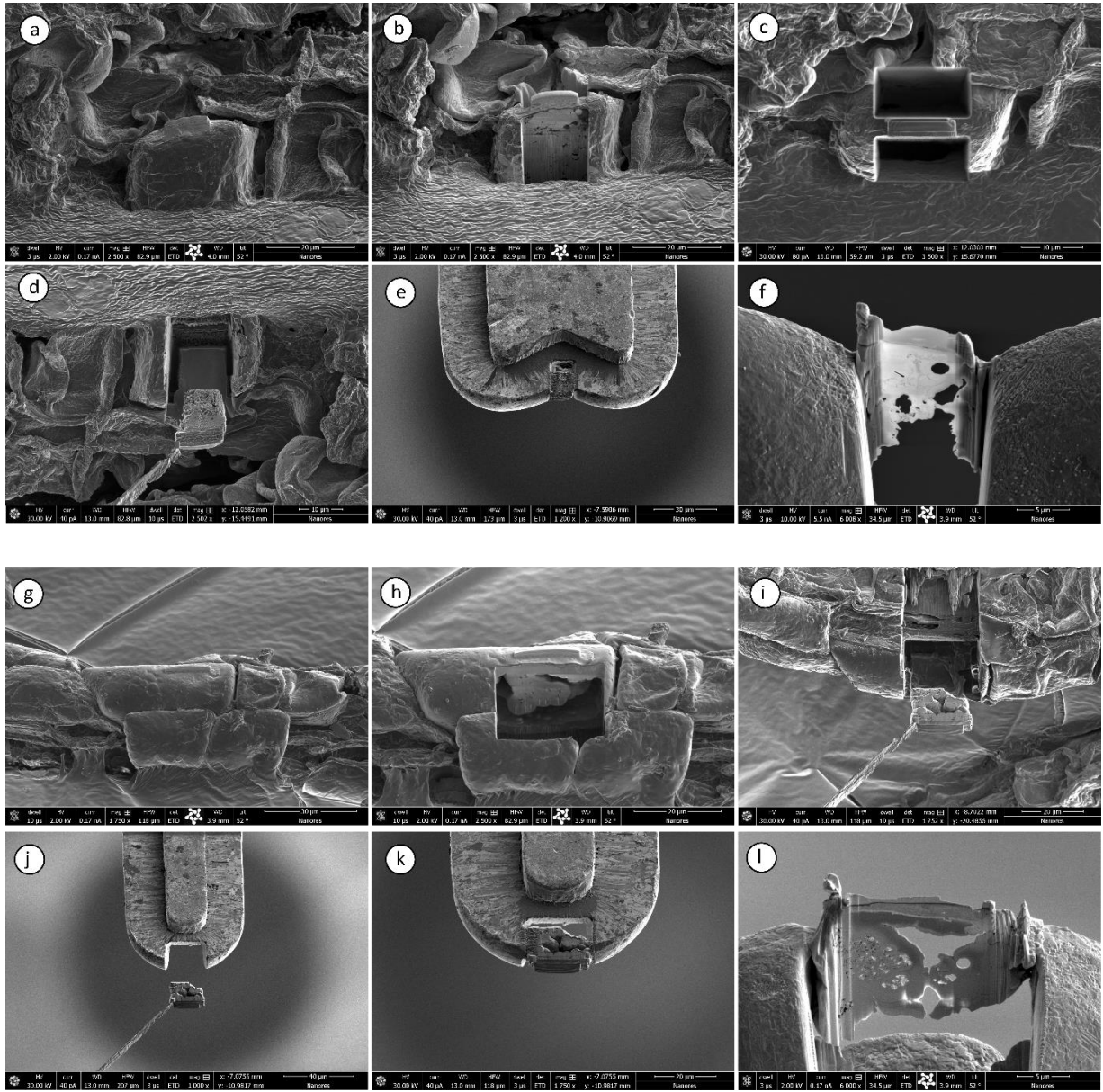

Fig. S3. Focused Ion Beam (FIB) assisted preparation of BB1 (a-f) and BM1 (g-l) thin sections. (a,g) ZnS mineralized root cell with 10 mm long Pt protection layer (52° tilt view); (b,h) FIB removal of material at both sides of the Pt layer ; (c) milled trenches viewed from above (0 tilt); (d, i)  in-situ lift-out; (e,j,k)  fixation on the TEM transfer grid (f,l) thin section (lamella) after thinning to electron transparency.

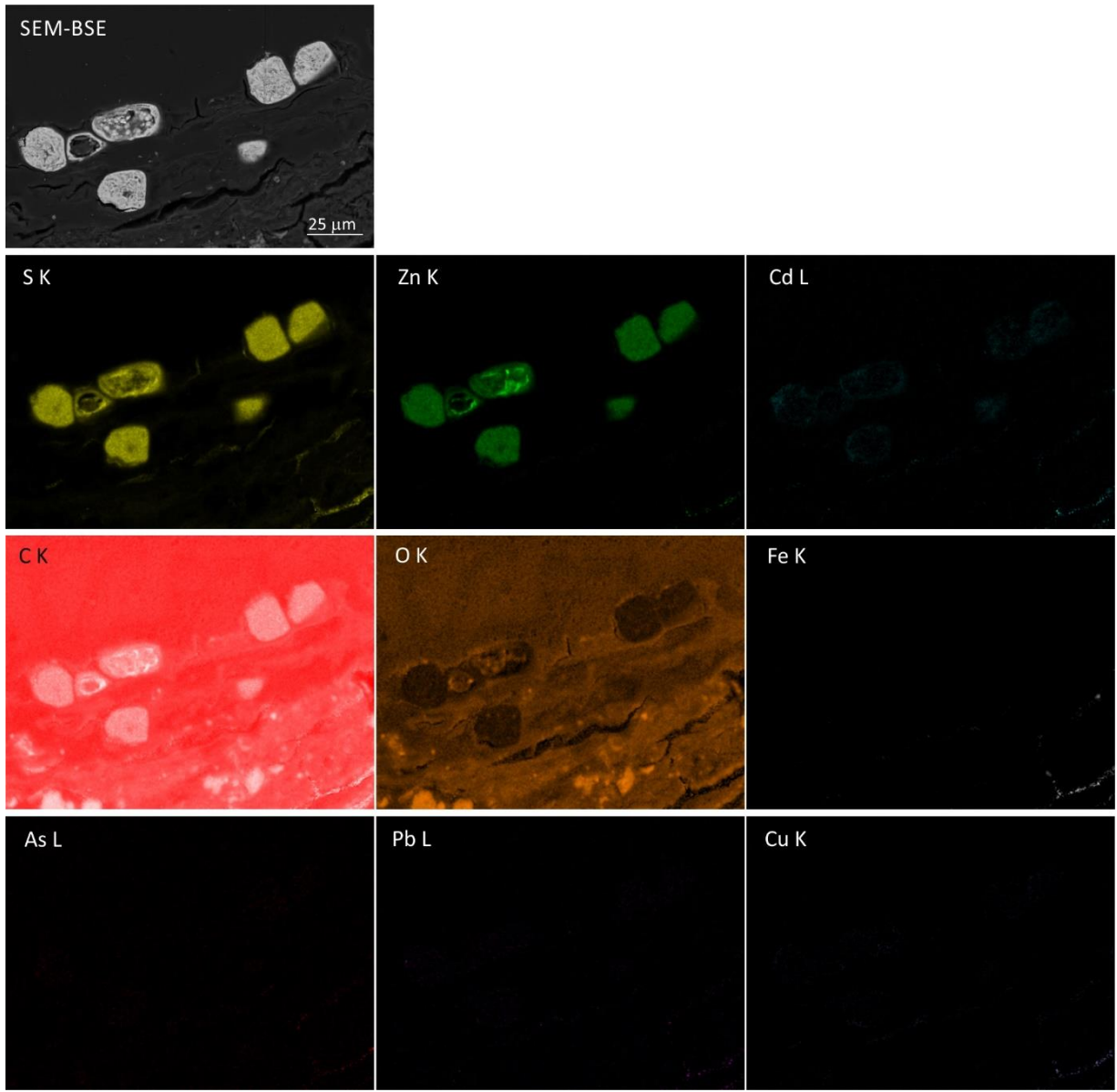

Fig. S4. SEM-BSE image of ZnS plant cell infillings and the corresponding SEM-EDS chemical (atomic %) distribution maps for S, Zn, Cd, C, O, Fe, As, Pb, and Cu (BM1 19-20cm). Low amounts of Cd and the absence of Fe, As, Pb, Cu are indicated for the ZnS precipitates. The decline in C and O at the sulfide site is visible.

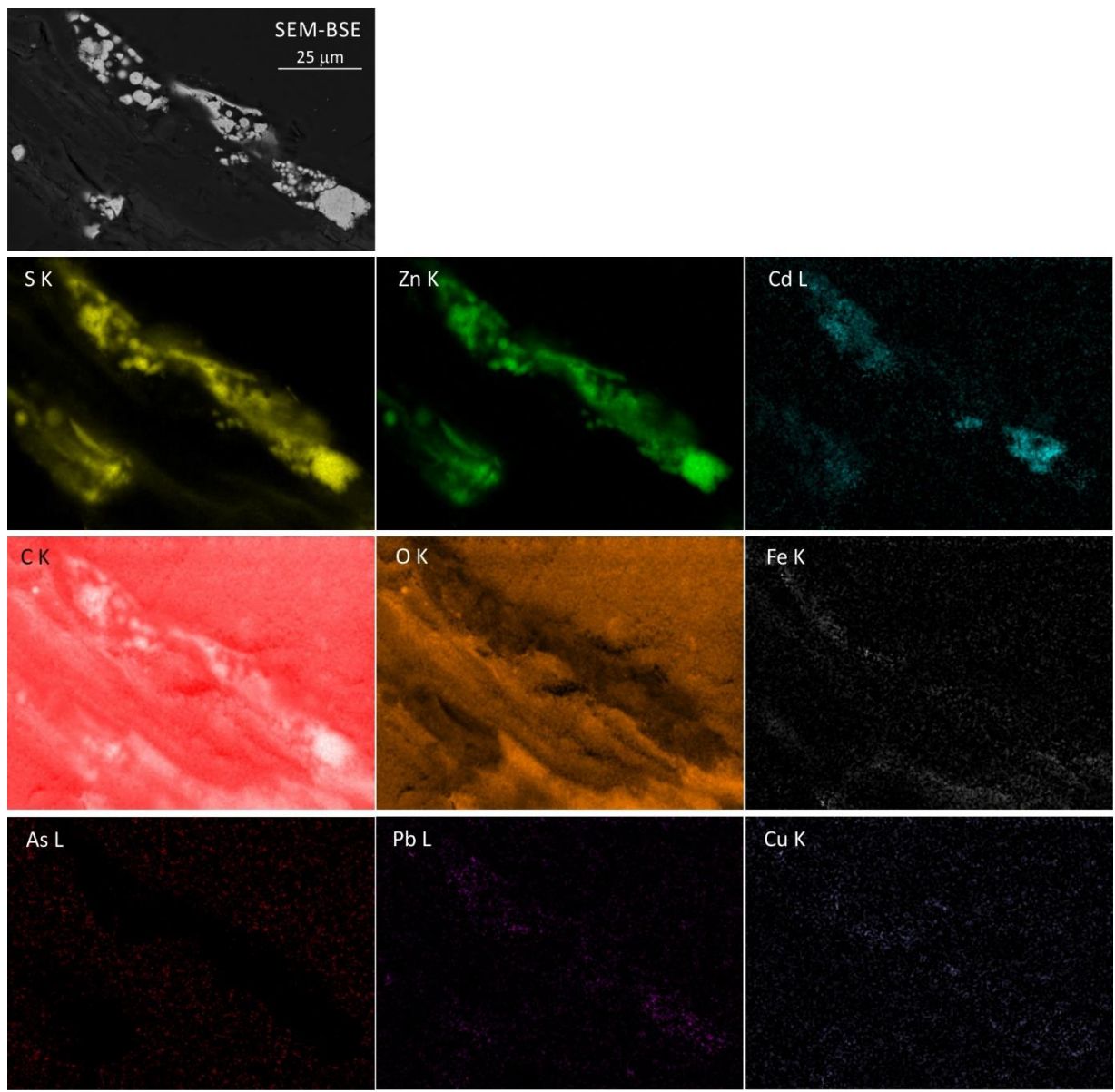

Fig. S5. SEM-BSE image of spherical ZnS aggregates inside plant tissues (upper right corner is an epoxy resin) and the corresponding SEM-EDS chemical (atomic %) distribution maps for S, Zn, Cd, C, O, Fe, As, Pb, and Cu (BM2 17-18cm). Varied amounts of Cd and traces of Pb are associated with the ZnS spheroids. Fe is associated with organic matter. The decline in C, O, and As (the *As L* peak overlaps with *Mg K*; therefore, the *As L* map shows As and Mg or Mg distribution) at the sulfides site is visible.

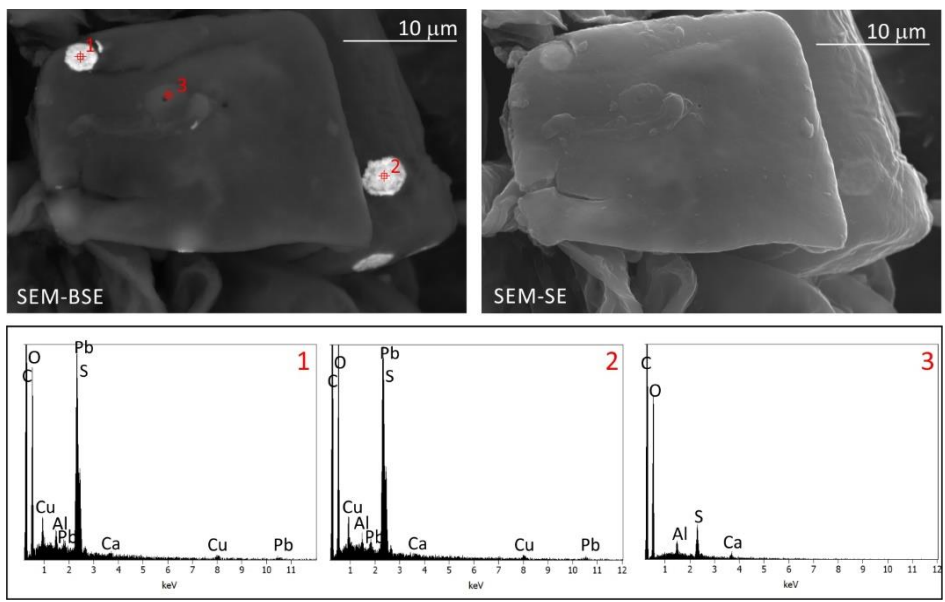

Fig. S6. BSE and SE images are set together to show the PbS precipitates to reside inside a plant cell. EDS spectra indicate the Cu association with PbS (spectrum 1 and 2). Al and Ca are associated with organic matter (compare spectra 1 and 2 with spectrum 3).

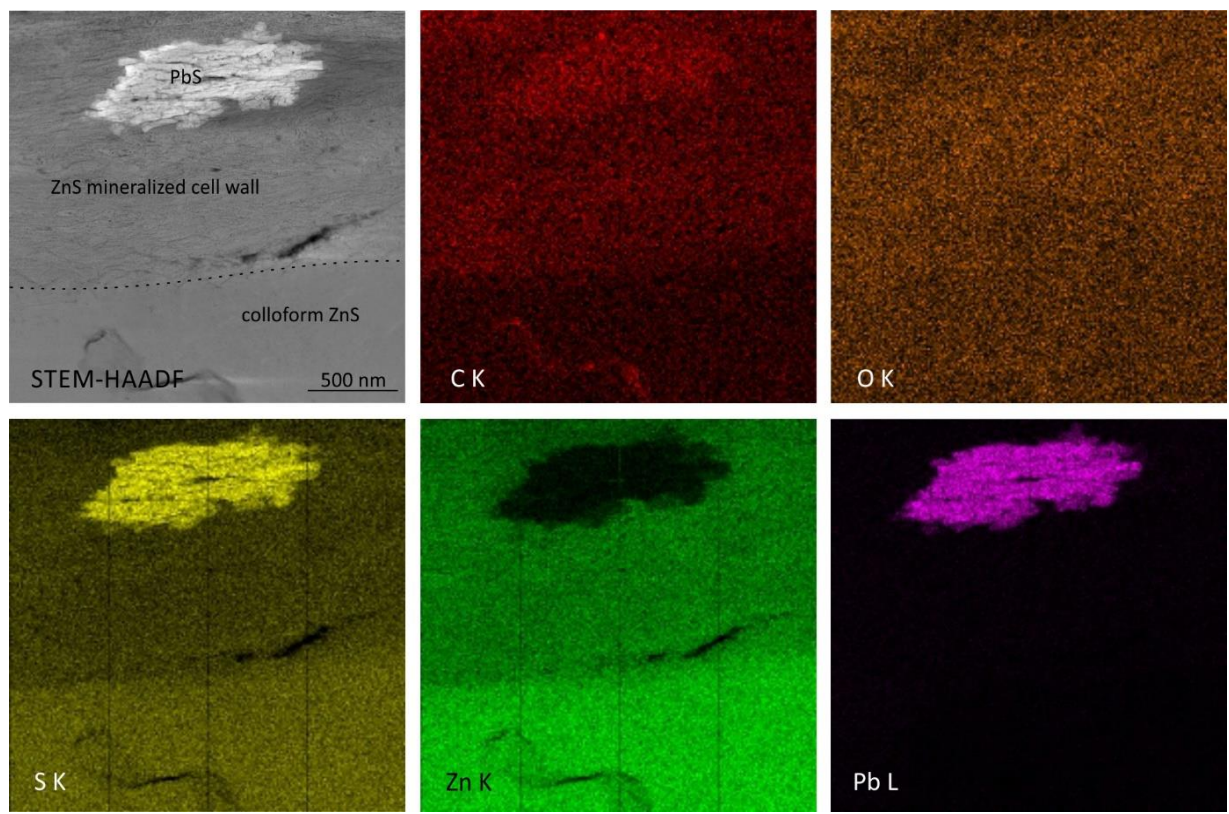

Fig. S7. STEM-HAADF image of a fragment of FIB-extracted ZnS mineralized plant cell with PbS inclusion and the corresponding STEM-EDS chemical (atomic %) distribution maps for S, Zn, Pb, C, and O (BM1, 18-19 cm).